\documentclass[11pt,a4paper]{article}
\pdfoutput=1
\usepackage[utf8]{inputenc}
\usepackage{jheppub}
\usepackage{braket,amsmath,amsthm,amssymb,amsbsy,amsfonts,mathrsfs,enumerate,float,wrapfig,mathtools,graphicx,graphbox,framed,tcolorbox,bbm,subfigure}
\usepackage{todonotes}
\newcommand{\lt}{\left}
\newcommand{\rt}{\right}
\usepackage{tikz,tkz-euclide}
\usetikzlibrary{patterns}
\tikzset{every picture/.style={line width=0.75pt}} 
\usetikzlibrary{positioning}
\usetikzlibrary{arrows, arrows.meta}
\usetikzlibrary{decorations, decorations.markings, decorations.pathmorphing}

\title{Topological Vertex for Symmetric matter}

\author[a]{Sung-Soo Kim,}
\author[b]{Xiaobin Li,}
\author[c]{Futoshi Yagi,}
\author[d]{Rui-Dong Zhu}
 \affiliation[a]{School of Physics, University of Electronic Science and Technology of China, \\
 No.2006 Xiyuan Ave, West Hi-Tech Zone, Chengdu, Sichuan 611731, China}
 \affiliation[b]{School of Mathematics, Southwest Jiaotong University,\\
 West zone, High-tech district, Chengdu, Sichuan 611756, China}
 \affiliation[c]{School of Science, Huzhou University,
 759 Erhuan East Road, Huzhou, Zhejiang, 313000, China}
 \affiliation[d]{Institute for Advanced Study \& School of Physical Science and Technology, Soochow University, Suzhou 215006, China}

\emailAdd{sungsoo.kim@uestc.edu.cn}
\emailAdd{lixiaobin@home.swjtu.edu.cn}
\emailAdd{60714@zjhu.edu.cn}
\emailAdd{rdzhu@suda.edu.cn}

\abstract{We propose a novel topological vertex formalism for 5d $\mathcal{N}=1$ SU($N$) gauge theory with a hypermultiplet in the symmetric tensor representation, whose Type IIB brane construction involves an NS5-brane attached to an O7$^+$-plane. Inspired by the identification $\mathrm{O7}^+\sim \mathbb{Z}_2 + 4 \mathrm{D7}$, we introduce two new types of vertices: the $\mathbb{Z}_2$-vertex, which implements the $\mathbb{Z}_2$ orbifold action, and the FD-vertex, which encodes the monodromy cut induced by the O7$^+$-plane. This formalism generalizes the framework presented in \cite{Kim:2024ufq} and establishes a systematic method for computing partition functions for 5-brane configurations that incorporate an O7$^+$-plane. The resulting partition functions are expressed as sums over Young diagrams, providing a powerful computational tool for studying such gauge theories.}

\begin{document}
\maketitle

\section{Introduction}\label{sec:intro}

The topological vertex formalism~\cite{Aganagic:2003db, Iqbal:2007ii, Awata:2008ed} is a powerful and systematic computational technique for studying non-perturbative aspects of supersymmetric gauge theories in five dimensions~(5d). Originally conceived for topological string theory on toric Calabi-Yau threefolds~(CY3), the formalism enables the exact calculations of partition functions and BPS spectra by recasting gauge theory dynamics in terms of string theory brane configurations. Specifically, 5d $\mathcal{N}=1$ supersymmetric gauge theories can be engineered using 5-brane webs in Type IIB string theory~\cite{Aharony:1997ju, Aharony:1997bh, Benini:2009gi}, where the rich interplay between geometry and gauge theory is captured elegantly by the combinatorial rules of the topological vertex. When these theories are compactified on a circle, they reveal intricate structures, such as instanton corrections and enhanced global symmetries, all of which can be systematically computed within the topological vertex framework. Thus, the topological vertex formalism not only provides a bridge between string theory and gauge theory, but also offers deep insight into the moduli spaces and dualities intrinsic to 5d $\mathcal{N}=1$ superconformal theories.

Recent developments have extended the topological vertex approach beyond toric CY3 geometries, incorporating 5-brane webs with orientifolds such as O5- and ON-planes~\cite{Kim:2017jqn, Bourgine:2017rik, Kim:2022dbr}. These advances establish the topological vertex as a versatile tool for computing Nekrasov instanton partition functions for 5d gauge theories on the Omega background~\cite{Nekrasov:2002qd, Nekrasov:2003rj}. Notable examples include Sp($N$)~\cite{Kim:2017jqn, Li:2021rqr}, SO($N$)~\cite{Hayashi:2020hhb, Kim:2021cua}, and  $G_2$~\cite{Hayashi:2018bkd} gauge theories, D-type quiver gauge theories~\cite{Bourgine:2017rik, Kim:2022dbr}, and also for six-dimensional theories with/without $\mathbb{Z}_2$ twist~\cite{Kim:2021cua, Kim:2020npz}.

A 5-brane web with an O5-plane also admits SO gauge theories with matter in the spinor representations~\cite{Zafrir:2015ftn, Zafrir:2016jpu}. Motivated by this setup, a new type of vertex, called the O-vertex~\cite{Hayashi:2020hhb}, is introduced, and plays a central role in constructing a topological vertex formalism for 5-brane web diagrams with an O5-plane. The O-vertex, though artificial as it is introduced at the intersection of the O5-plane and 5-branes, enables one to compute the (unrefined) partition function for SO gauge theories as a sum over Young diagrams~\cite{Hayashi:2020hhb, Nawata:2021dlk}, which is consistent with results from ADHM construction \cite{Nekrasov:2004vw}.

Like 5-brane web with an O5-plane, a 5-brane web with an O7-plane is also non-toric. Although both SO and Sp gauge theories can be realized with either an O7- or O5-plane~\cite{Bergman:2015dpa, Zafrir:2015ftn}, the pertinent orientifold projections differ: The O7-plane imposes a $\mathbb{Z}_2$ symmetry corresponding to a $\pi$-rotation of the 5-brane plane, whereas the O5-plane implements a $\mathbb{Z}_2$ reflection across a fixed plane. This difference leads to different brane descriptions for yet another gauge theory. A 5-brane web with an O7$^\pm$-plane where an NS5-brane is attached yields SU gauge theories with a hypermultiplet in the symmetric / antisymmetric tensor representation, respectively. These hypermultiplets cannot be achieved using O5-planes. However, the corresponding topological vertex formalism for these 5-brane webs with O7-planes had not yet been fully developed.

Recently, the authors have developed topological vertex formalism with an O7$^+$-plane~\cite{Kim:2024ufq}, exploiting the insight that $\mathrm{O7}^+\sim \mathbb{Z}_2+4\mathrm{D7}$. While the resulting partition functions agree with those obtained using the O-vertex formulations, this approach does not extend to 5-brane configurations with an O7$^+$-plane where an NS5-brane is attached. In particular, it is not applicable to SU gauge theories with a symmetric hypermultiplet. In this work, we generalize the construction of~\cite{Kim:2024ufq}, presenting a formulation accommodates 5-brane webs with an O7$^+$-plane irrespective of NS5-brane attachments.

Our proposal for a new topological vertex with an O7$^+$-plane builds upon and extends the identification $\mathrm{O7}^+\sim \mathbb{Z}_2+4\mathrm{D7}$, which is a variant of the so-called {\it freezing} prescription, $\mathrm{O7}^+\sim \mathrm{O7}^-+8\mathrm{D7}$, where the masses of eight D7-branes are precisely tuned at the position of an O7$^-$-plane, allowing a pairwise $\pi\,i$ shift~\cite{Hayashi:2023boy, Kim:2023qwh, Kim:2024vci}. The freezing technique has yielded effective computations in theories with O7$^+$-planes. For instance, the non-Lagrangian theory, local $\mathbb{P}^2+1\mathbf{Adj}$ \cite{Bhardwaj:2019jtr} requires an O7$^+$-plane for its brane realization~\cite{Kim:2020hhh}, and its Seiberg-Witten curve~\cite{Hayashi:2023boy} and superconformal index~\cite{Kim:2023qwh} can be obtained by applying the freezing method. Moreover, the BPS spectra of theories with an O7$^+$-plane are shown to be closely related to those with an O7$^-$-plane~\cite{Kim:2024vci}.

The main novelty in our topological vertex prescription for the O7$^+$-plane consists of two essential ingredients: 
(i) the explicit construction utilizing the $\mathbb{Z}_2$ orbifold fixed point attached to an NS-brane, and 
(ii) the introduction of a new vertex associated with the intersection between a ($p,1$) 5-brane and the monodromy cut of the O7$^+$-plane. Taken together, our formalism establishes a systematic framework for computing partition functions for 5d $\mathcal{N}=1$ supersymmetric SU($N)_k$ gauge theories with a symmetric hypermultiplet, as well as SO($2N$) gauge theories.

There are other topological vertex formulations in the
presence of orientifold planes in CY3. For instance, the real
topological string~\cite{Walcher:2007qp, Krefl:2009md, Krefl:2009mw} is constructed for toric CY 3-folds with $\mathbb{Z}_2$ orientifolds. In contrast, our formulation of the topological vertex with orientifolds is specifically designed for 5-brane web diagrams containing an O7-plane (or an O5-plane in earlier works), and thus differs from the real topological vertex. The distinctions between the two approaches are discussed in Appendix~\ref{sec:App-RTV}.

The paper is organized as follows. In section~\ref{sec:reformulation}, we recast the Nekrasov instanton partition function for 5d $\mathcal{N}=1$ SU$(N)_k$ gauge theory at Chern-Simons level $k$ with a symmetric hypermultiplet into a form suitable for our new topological vertex formalism. Section~\ref{sec:O7-vertex} introduces our proposal for the topological vertex formalism for the O7$^+$-plane, where we construct two novel vertices--the $\mathbb{Z}_2$-vertex and the FD-vertex--that together enable the computation of the partition function for SU$(N)_k$ gauge theory with a symmetric hypermultiplet. In section~\ref{sec:conclusion}, we summarize our results and discuss potential future directions. The notation, conventions, and identities used in the paper are detailed in Appendix~\ref{sec:App}. We compare the differences between our topological vertex with an O7-plane and the real topological vertex \cite{Walcher:2007qp, Krefl:2009md, Krefl:2009mw} in Appendix~\ref{sec:App-RTV}.


\bigskip
\section{Reformulation of Nekrasov partition function} \label{sec:reformulation}

In this section, we rewrite the Nekrasov partition function for five-dimensional (5d) $\mathcal{N}=1$ SU$(N)_k$ gauge theory at Chern-Simons level $k$ with a hypermultiplet in the symmetric tensor representation ($\mathbf{Sym}$) into an expression that is useful for the discussion on the topological vertex formalism proposed in section \ref{sec:O7-vertex}.

\subsection{Rewriting Nekrasov partition function}\label{sec:rewriting}
The Nekrasov partition function for 5d $\mathcal{N}=1$ SU$(N)_k + 1 \, \mathbf{Sym}$ is given~\cite{Chen:2023smd} by a sum over the Young diagrams $\vec\lambda$ as
\begin{align}\label{eq:Nek-SU+Sym-original}
Z_{\mathrm{SU}(N)_k+1\mathbf{Sym}}(\vec{a},m) 
= \sum_{\vec{\lambda}} Q_I^{|\vec{\lambda}|}
z_{\mathrm{vec}}(\vec{a},\vec{\lambda}) z_{\mathrm{sym}}(\vec{a},m,\vec{\lambda})\ ,
\end{align}
in the unrefined limit
\begin{align}
\epsilon_1 = - \epsilon_2 := \hbar.
\end{align}
Here, $Q_I$ denotes the instanton factor while $z_{\mathrm{vec}}$ and $z_{\mathrm{sym}}$ represent the contributions from the vector multiplet and the symmetric hypermultiplet, respectively.

The contribution from the vector multiplet takes the form,  \begin{align}\label{eq:Zvec}
z_{\mathrm{vec}}(\vec{a},\vec{\lambda},k)
= (-1)^{|\vec{\lambda}|N} \prod_{s=1}^N \prod_{x \in \lambda_s} 
e^{k \phi_s(x)} \prod_{t=1}^N \frac{1}{\mathrm{sh}^2 \big(a_s - a_t - \hbar (a_{\lambda_s(x)} + l_{\lambda_t}(x) + 1)\big)}\ ,
\end{align}
where 
\begin{align}\label{eq:def-phi}
x= (i,j),
\qquad
\phi_s(x) &:= a_s +(i-j) \hbar \ , 
\end{align}
and 
\begin{align}
\mathrm{sh}(z) =e^{z/2}- e^{-z/2} \ , 
\qquad 
a_{\lambda(x)} = (\lambda)_{i} - j \ ,
\qquad
l_{\lambda(x)} = (\lambda^T)_{j} - i \ .
\end{align}
It is convenient to introduce the new Nekrasov factor%
\footnote{Although this paper is not the first one to introduce this factor, we use the word ``new'' Nekrasov factor to distinguish it from the original Nekrasov factor \eqref{def:Nekra} and to use consistent terminology with \cite{Kim:2024ufq}.}
,
\begin{align}
\tilde{n}_{\lambda\mu}(a;\hbar) 
:=& \prod_{(i,j)\in \lambda}
\mathrm{sh}\Big(a + \hbar \big(1-i-j+(\lambda)_i+(\mu^T)_j \big)\Big)
\cr
& \times
\prod_{(i',j')\in \mu}
\mathrm{sh}\Big(a + \hbar \big(i'+j'-1-(\lambda^T)_{j'}-(\mu)_{i'} \big)\Big)\ , 
\label{eq:def-nekt}
\end{align}
so that the vector contribution \eqref{eq:Zvec} is re-expressed in terms of the new Nekrasov factor 
\begin{align}\label{eq:vec-cont}
z_{\mathrm{vec}}(\vec{a},\vec{\lambda}, k)
= \prod_{s=1}^N e^{k \big(|\lambda_s| a_s - \frac{\kappa(\lambda_s)}{2}\hbar \big)} \times 
\prod_{s,t=1}^N 
\frac{1}{\tilde{n}_{\lambda_s\lambda_t} (a_s-a_t)}\ ,
\end{align}
where $\kappa(\lambda)=2\sum_{(i,j)\in\lambda}(j-i)$. (See also Appendix \ref{sec:App} for various identities that $\tilde{n}_{\lambda\mu}$ satisfies.)

The contribution from the hypermultiplet of mass $m$ in the symmetric tensor representation takes the form, 
\begin{align}\label{eq:sym-cont}
z_{\mathrm{sym}}(\vec{a},m,\vec{\lambda})
=&~  (-1)^{[\frac{|\vec{\lambda}|}{2}]} 
\prod_{s=1}^N 
\prod_{x \in \lambda_s} 
\mathrm{sh} \big(2 \phi_s(x) \!+\! m \pm \hbar \big)
\prod_{s,t=1}^N
\prod_{x \in \lambda_s} 
\mathrm{sh} \big(\phi_s(x) + a_t\! +\! m\big)
\cr
& \times
\prod_{1 \le s \le t \le N} \prod_{\substack{x \in \lambda_s, y \in \lambda_t \\{x<y}}} \Delta_{s,t}(x,y)\ , 
\end{align}
where we used the shorthand notations,
$\mathrm{sh}(x\pm y) := \mathrm{sh}(x+y)\cdot\mathrm{sh}(x-y)$ and 
\begin{align}\label{eq:def-phi-delta}
\Delta_{s,t}(x,y) &:= \frac{\mathrm{sh}(\phi_s(x)+\phi_t(y)+m\pm \hbar)}{\mathrm{sh}^2(\phi_s(x)+\phi_t(y)+m)}\ .
\end{align}
Here, we have defined the total ordering on the boxes of the Young diagrams $\lambda_s \ni x=(i, j)$ and $\lambda_t\ni y=(i' , j')$ as 
\begin{align}
x < y 
\qquad \text{if} \quad
\left\{
\begin{array}{l}
s<t \\
s = t,~ i < i' \\
s = t,~ i = i',~ j < j' 
\end{array}
\right. .
\end{align}

Similar to the vector contribution, the hypermultiplet contribution can also be reformulated in terms of the new Nekrasov factors. To achieve this, we implement the following modifications. First, we utilize the symmetry property $\Delta_{s,t}(x,y)=\Delta_{t,s}(y,x)$ to recast the second line in \eqref{eq:sym-cont} as%
\footnote{
We have also included the sign factor $(-1)^{[\frac{|\vec{\lambda}|}{2}]}$ in the first line in \eqref{eq:sym-cont} and have assumed that it can be absorbed into the sign ambiguity related to the square root. Although we have not justified this assumption at this stage, it turns out to be consistent with the computation in the later section.
}
\begin{align}\label{eq:rewrite-Delta}
 (-1)^{[\frac{|\vec{\lambda}|}{2}]} \prod_{1 \le s \le t \le N} \prod_{\substack{x \in \lambda_s, y \in \lambda_t \\ {x<y}}} \Delta_{s,t}(x,y)
 = \prod_{s,t=1}^N \prod_{\substack{x \in \lambda_s \\ y \in \lambda_t}} \Delta_{s,t}(x,y)^{\frac12}
\times 
\prod_{s=1}^N \prod_{x \in \lambda_s } \Delta_{s,s}(x,x)^{-\frac12}\ .
\end{align}
As shown in Appendix \ref{sec:App}, the new Nekrasov factor satisfies the following identity or \eqref{eq:key-id-n}:
\begin{align}\label{eq:key-id-main}
\prod_{\substack{(i,j) \in \lambda \\{(i',j') \in \mu}}} 
\frac{\mathrm{sh} \big( a+(i-j+i'-j' \pm 1)\hbar \big)}{\mathrm{sh}^2 \big( a+(i-j+i'-j')\hbar \big)}
= 
\frac{\tilde{n}_{\lambda^T\mu}(a)
}{\displaystyle
\prod_{(i,j) \in \lambda} \mathrm{sh} \big( a+(i-j)\hbar \big)
\prod_{(i',j') \in \mu} \mathrm{sh} \big( a+(i'-j')\hbar \big)}.
\end{align}
This helps us to rewrite \eqref{eq:rewrite-Delta} in terms of the new Nekrasov factor. By replacing $a \to a_s + a_t + m$, in~\eqref{eq:key-id-main}, we can see that the first factor on the right-hand side in \eqref{eq:rewrite-Delta} is given by
\begin{align}\label{eq:sh-to-ntilde}
& \prod_{s,t=1}^N \prod_{\substack{x \in \lambda_s \\ y \in \lambda_t}} \Delta_{s,t}(x,y)^{\frac12}
= 
\prod_{s,t=1}^N
\frac{\tilde{n}^{\frac12}_{\lambda_s^T \lambda_t}(a_s + a_t + m)
}{\displaystyle
\prod_{x \in \lambda_s} \mathrm{sh}( \phi_s(x) + a_t + m )}.
\end{align}
From \eqref{eq:sym-cont}, \eqref{eq:def-phi-delta}, \eqref{eq:rewrite-Delta}, and \eqref{eq:sh-to-ntilde}, we find that the symmetric hypermultiplet contribution is written as
\begin{align}\label{eq:z_sym_ntilde}
z_{\mathrm{sym}}(\vec{a},m,\vec{\lambda})
= & 
\prod_{s=1}^N 
\prod_{x \in \lambda_s} 
\mathrm{sh}^{\frac12} (2 \phi_s(x) + m \pm \hbar ) \,
\mathrm{sh}(2\phi_s(x)+m)
\times\!\!
\prod_{s,t=1}^N
\tilde{n}^{\frac12}_{\lambda_s^T \lambda_t}(a_s + a_t + m).
\end{align}

Next, we introduce four%
\footnote{As can be seen from \eqref{eq:sym-final}, they could have been more precisely denoted as ``half'' of ``eight'' virtual hypermultiplets, whose masses are given by $\pm \mathsf{m}_f$ ($f=1,2,3,4$). However, in this paper, we denote them as ``four'' for simplicity until the end of section \ref{subsec:fcon}, where we revisit this point.}
{\it virtual} hypermultiplets,
whose masses $\sf{m}_f$ are appropriately tuned according to the freezing procedure described in~\cite{Kim:2024ufq}, 
\begin{align}\label{eq:tuned-mass}
\mathsf{m}_f = \left\{ 0, ~\frac{\hbar}{2},~\pi i,~ \frac{\hbar}{2}+\pi i \right\}  . 
\end{align}
These hypermultiplets are considered ``frozen'' in the sense that their masses are fixed to specific values, thereby enabling the desired reformulation of the contribution of the symmetric hypermultiplet as 
\begin{align}\label{eq:sym-final}
z_{\mathrm{sym}}(\vec{a},m,\vec{\lambda})
= & 
\prod_{s=1}^N \prod_{f=1}^4
\prod_{x \in \lambda_s} 
\mathrm{sh}^{\frac12} (\phi_s(x) + \frac12 m \pm \mathsf{m}_f) \,
\times
\prod_{s,t=1}^N 
\tilde{n}^{\frac12}_{\lambda_s^T \lambda_t}(a_s + a_t + m)
\cr
= & 
\prod_{s=1}^N \prod_{f=1}^4
\tilde{n}^{\frac12}_{\lambda_s^T \varnothing}(a_s + \frac12 m \pm \mathsf{m}_f) \,
\times
\prod_{s,t=1}^N 
\tilde{n}^{\frac12}_{\lambda_s^T \lambda_t}(a_s + a_t + m)\ ,
\end{align}
where in the first equality, we used the identity
\begin{align}
\mathrm{sh} (2x) =- i \, \mathrm{sh} (x) \, \mathrm{sh} (x+\pi i) = i \, \mathrm{sh} (x) \, \mathrm{sh} (x-\pi i)\ ,
\end{align}
while in the second equality, we used
\begin{align}
\tilde{n}_{\lambda^T \varnothing}(a) = \prod_{(i,j) \in \lambda} \mathrm{sh} \big(a + \hbar (i-j) \big) \ , 
\end{align}
which follows from the definition of the new Nekrasov factor, \eqref{eq:def-nekt}. Also, see \eqref{eq:Nek-empty}.

Combined with the contribution from the vector multiplet~\eqref{eq:vec-cont}, we finally find that the Nekrasov partition function \eqref{eq:Nek-SU+Sym-original} for 5d SU($N$)$_k$ gauge theory with a hypermultiplet in the symmetric tensor representation is given by
\begin{align}\label{eq:Nek-SU+Sym-final}
Z_{\mathrm{SU}(N)_k+1\mathbf{Sym}}(\vec{a},m) 
= & \sum_{\vec{\lambda}} 
Q_I^{|\vec{\lambda}|}
\prod_{s=1}^N 
e^{k (|\lambda_s| a_s - \frac{\kappa(\lambda_s)}{2} \hbar )} 
\times \prod_{s,t=1}^N 
\frac{\tilde{n}^{\frac12}_{\lambda_s^{T}\lambda_t}(a_s + a_t + m)}{\tilde{n}_{\lambda_s\lambda_t} (a_s-a_t)}
\cr
& \qquad 
\times
\prod_{t=1}^{N} \prod_{f=1}^4  
\tilde{n}^{\frac12}_{\varnothing \lambda_t}(a_t + \frac12m \pm  \mathsf{m}_f)\ , 
\end{align}
where $\mathsf{m}_f$ are frozen as given in \eqref{eq:tuned-mass}.

One advantage of expressing the partition function in terms of the new Nekrasov factor $\tilde{n}_{\mu\nu}$ is that the structure of 4d limit is maintained. In other words, one can simply replace the 5d Nekrasov factors \eqref{eq:def-nekt} by 4d Nekrasov factors 
\begin{align}
\tilde{n}^{4d}_{\lambda\mu}(a;\hbar) 
:=& \prod_{(i,j)\in \lambda}
\Big(a + \hbar \big(1-i-j+(\lambda)_i+(\mu^T)_j \big)\Big)
\cr
& \times
\prod_{(i',j')\in \mu}
\Big(a + \hbar \big(i'+j'-1-(\lambda^T)_{j'}-(\mu)_{i'} \big)\Big)\ .
\label{eq:def-nek4d}
\end{align}
In addition, the factor $e^{k (|\lambda_s| a_s - \frac{\kappa(\lambda_s)}{2} \hbar )} $, which depends on Chern-Simons level $k$, reduces to 1 in the 4d limit. 
Therefore, we can readily see that the 4d limit of our result is consistent with \cite{Marino:2004cn}.

It is also worth noting that \eqref{eq:Nek-SU+Sym-final} is rewritten as
\begin{align}\label{eq:gen-KN}
Z_{\mathrm{SU}(N)_k+1\mathbf{Sym}}(\vec{a},m) 
= &
\sum_{\vec{\lambda}} Q_I^{|\vec{\lambda}|}
\biggl( 
z_{\mathrm{vec}}(\vec{a},\vec{\lambda}, k) \, 
z_{\mathrm{vec}}(-\vec{a}, \vec{\lambda}^T , -k) \,
z_{\text{bif}} (\vec{a}, - \vec{a}, m, \vec{\lambda}^{T}, \vec{\lambda})
\cr
& \quad
\prod_{f=1}^{4}  
z_{\text{fund}} (\vec{a}, \frac12 m + \mathsf{m}_{f} , \vec{\lambda})
\,
z_{\text{fund}} (-\vec{a}, - \frac12 m + \mathsf{m}_{f}, \vec{\lambda}^T)
\biggr)^{\frac12} \ , 
\end{align}
where we denote the contributions from the hypermultiplets in the bi-fundamental representation and the fundamental representation, respectively, as
\begin{align}\label{eq:def-bif-fund}
z_{\text{bif}} (\vec{a}^{(1)}, \vec{a}^{(2)}, m, \vec{\lambda}^{(1)}, \vec{\lambda}^{(2)})
&:= \prod_{s,t=1}^N \tilde{n}_{\lambda_s^{(1)} \lambda_t^{(2)}} (a_s^{(1)} - a_t^{(2)} + m ) , \
\cr
z_{\text{fund}} (\vec{a}, m, \vec{\lambda})
&:= \prod_{s=1}^N \tilde{n}_{\varnothing \lambda_s} (a_s + m ) , \
\end{align} 
while the vector contribution is given in \eqref{eq:vec-cont}. 
The mass parameters $\mathsf{m}_f$ are identical to the ones given in \eqref{eq:tuned-mass}.
One may observe that the summand in \eqref{eq:gen-KN} is the square root of the summand of the Nekrasov partition function for $\mathrm{SU}(N)_{k^{(1)}} \times \mathrm{SU}(N)_{k^{(2)}}$ gauge theory with a bi-fundamental hypermultiplet and four fundamental hypermultiplets for each gauge group
\begin{align}
&z_{\mathrm{vec}}(\vec{a}^{(1)},\vec{\lambda}^{(1)}, k^{(1)}) \, 
z_{\mathrm{vec}}(\vec{a}^{(2)}, \vec{\lambda}^{(2)} , k^{(2)}) \,
z_{\text{bif}} (\vec{a}^{(1)}, \vec{a}^{(2)}, m, \vec{\lambda}^{(1)}, \vec{\lambda}^{(2)})
\cr
& \qquad \times \prod_{f=1}^4 
z_{\text{fund}} (\vec{a}^{(1)}, m_{f}^{(1)} , \vec{\lambda}^{(2)})
\times \prod_{f=1}^4  
z_{\text{fund}} (\vec{a}^{(2)}, m_{f}^{(2)} , \vec{\lambda}^{(2)}) \ ,
\end{align}
subject to specific tunings of Chern-Simons level, Young diagrams, Coulomb moduli, and masses as
\begin{align}
&k^{(1)} = - k^{(2)} :=k, \quad
\lambda_i^{(1)} = \left( \lambda_i^{(2)} \right)^T, \quad 
a_i^{(1)} = - a_i^{(2)} := a_i, \quad  
\cr
&m_f^{(1)} = - m_f^{(2)} + \frac{\hbar}{2} = \frac12 m + \mathsf{m}_f\ .
\end{align}
This observation can be regarded as a generalization of the observation on the partition functions between SU($N$) gauge theory and SO($2N$) gauge theory found in \cite{Lee:2024jae}.

\subsection{Consistency check}\label{subsec:cons-ADHM}
As a consistency check, we consider the following Higgsing, where SU($2N)+1\mathbf{Sym}$ admits a Higgs branch to SO$(2N)$ theory without any hypermultiplet. Here we explicitly show that our proposal indeed leads to SO$(2N)$ theory by tuning the Coulomb branch parameters and the mass of the symmetric hypermultiplet,
\begin{align}\label{eq:tuning-a-m}
a_s = - a_{2N+1-s}, \qquad m = 0 \ . 
\end{align}
Based on the partition function \eqref{eq:Nek-SU+Sym-final}, we perform the Higgsing on SU($2N)_k+1\mathbf{Sym}$ which reduces to SO($2N$). First, we tune the mass according to~\eqref{eq:tuning-a-m}: 
\begin{align}\label{eq:Nek-tuned}
Z_{\mathrm{SU}(2N)_k+1\mathbf{Sym}}
\xrightarrow{m=0} &  
\sum_{\lambda_1, \cdots, \lambda_{2N}} 
Q_I^{|\vec{\lambda}|}
\prod_{s=1}^{2N} 
e^{k (|\lambda_s| a_s - \frac{\kappa(\lambda_s) 
}{2} )} 
\prod_{s,t=1}^{2N} 
\frac{\tilde{n}^{\frac12}_{\lambda_s^{T}\lambda_t}(a_s + a_t)}{\tilde{n}_{\lambda_s\lambda_t} (a_s-a_t)}
\cr
& \qquad 
\times
\prod_{t=1}^{2N} \prod_{f=1}^4  
\tilde{n}^{\frac12}_{\varnothing \lambda_t}(a_t \pm \mathsf{m}_f)\ .
\end{align}
Implementing the Coulomb branch parameter as given in \eqref{eq:tuning-a-m}, when $t=2N+1-s$, we have the factors proportional to the delta functions, reducing the number of relevant Young diagrams by half,
\begin{align}
\prod_{s=1}^{2N} \tilde{n}^{\frac12}_{\lambda_s^{ T}\lambda_{2N+1-s}}(a_s+a_{2N+1-s}) 
= \prod_{s=1}^{2N} \tilde{n}^{\frac12}_{\lambda_s^{ T}\lambda_{2N+1-s}}(0) 
\propto \prod_{s=1}^{2N} \delta_{\lambda_s^{T}, \lambda_{2N+1-s} }\ .
\end{align}
Here, we used Eq.~(3.22) of \cite{Cheng:2018wll}. This suggests that instead of summing all the possible $2N$ Young diagrams $\lambda_{s}$ ($s=1,2,\cdots, 2N$), it is enough to take the summation over only $\lambda_{s=1,\cdots, N}$, while the remaining ones are tuned to be 
\begin{align}\label{eq:tuning-lambda}
\lambda_{2N+1-s} = \lambda_s^{T} \quad (s = 1, 2, \cdots, N)\ .
\end{align}
Under the Higgsing by tuning \eqref{eq:tuning-a-m} and \eqref{eq:tuning-lambda}, the partition function for SU($2N)_k+1\mathbf{Sym}$ \eqref{eq:Nek-SU+Sym-final} reduces to 
\begin{align}
Z_{\mathrm{SU}(2N)_k+1\mathbf{Sym}}\!
\xrightarrow{\rm Higgsing} \! \!
\sum_{\lambda_1, \cdots, \lambda_N} \!\!
(-Q_I^2)^{\sum_{s=1}^N |\lambda_s|} 
\left(
\frac{\displaystyle\prod_{t=1}^{2N} \prod_{f=1}^4
\tilde{n}_{\varnothing \lambda_t}(a_t \pm \mathsf{m}_f)}{\displaystyle\prod_{s=1}^{2N} \prod_{t=1}^{2N} \tilde{n}_{\lambda_s\lambda_t} (a_t-a_s)}
\right)^{\!\frac12}\!\! = Z_{\mathrm{SO}(2N)},
\end{align}
where we used \eqref{eq:sym-new-nek}.
This is exactly the expression for the SO($2N$) Nekrasov partition function~\cite{Lee:2024jae, Kim:2024ufq} up to the redefinition of the instanton factor.

\section{Topological vertex with \texorpdfstring{O7$^+$}{O7+}-plane}\label{sec:O7-vertex}

In this section, we propose a novel prescription for the unrefined topological vertex formalism applicable when a $(p,1)$ 5-brane is attached to an O7$^+$-plane. Using our prescription, we compute the partition function for the 5d $\mathcal{N}=1$ SU$(N)_k$ gauge theory at Chern-Simons level $k$ with a hypermultiplet in the symmetric tensor representation.

\subsection{Proposal of new prescription for topological vertex with \texorpdfstring{O7$^+$-plane}{O7+-plane}}\label{sec:proposal}
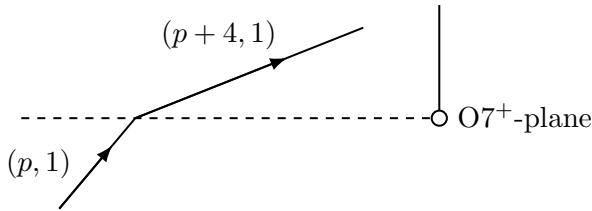
\begin{figure}[ht]
\centering
\begin{tikzpicture}[
    >={Latex},
    dot/.style = {circle, draw, fill=white, inner sep=0pt, minimum size=2pt}
]
\draw[dashed] (-3,0) -- (2.4,0);
\draw (2.5,0.1) -- (2.5,1.5);
\draw (2.5,0) circle (0.1cm);
\node[anchor=west] at (2.6,0) {O7$^{+}$\text{-plane}};
\draw (-2.5,-1.2) -- (-1.5,0);
\draw[->] (-2.5,-1.2) -- (-1.8,-0.36);
\node[anchor=east] at (-2.2,-0.6) {$(p,1)$};
\draw (-1.5,0) -- (1.5,1.2);
\draw[->] (-1.5,0) -- (0.5,0.8);
\node[anchor=east] at (0.5,1.1) {$(p+4,1)$};
\end{tikzpicture}
\caption{The change of the 5-brane charge from $(p,1)$ to $(p+4,1)$ after going across the monodromy cut created by an O7$^+$-plane, drawn in the fundamental region taken on the left side of the O7$^+$-plane.}
\label{fig:O7-monodromy}
\end{figure}

An O7$^+$-plane is equipped with the $\mathbb{Z}_2$ orbifold action that acts on spacetime in such a way as to rotate a 5-brane web diagram by $\pi$ radians. This means that on 5-brane webs, there are regions that are identified by this $\mathbb{Z}_2$ orbifold action. It is, hence, redundant to keep all the regions of the 5-brane and thus enough to keep only half of the given web. We call such a half web the fundamental region. We can choose the fundamental region to be, for example, the left half of the 5-brane web as shown in Figure~\ref{fig:O7-monodromy}. In addition, an O7$^+$-plane has a positive RR-charge, which causes the monodromy effect. We postulate that the corresponding monodromy cut extends in the horizontal direction from the position of the O7$^+$-plane. When $(p,q)$ 5-branes go across this monodromy cut, their $(p,q)$ charges change as in Figure~\ref{fig:O7-monodromy}. 

Given a 5-brane web diagram, one computes the partition function based on topological vertex formalism. We propose a topological vertex for a 5-brane web with an O7$^+$-plane. Analogously to the topological vertex formalism based on various web diagrams without an O-plane, it is natural to expect that the corresponding partition function is expressed as a sum over all the Young diagrams assigned to each edge of the 5-brane web. The summand is given by the product of the contributions from all edges and vertices. For trivalent vertices and edges connecting them, one should assign corresponding factors following standard prescriptions established for conventional toric web diagrams. For external edges that extend to infinity, they do not contribute to the partition function. 

With an O7$^+$-plane, topological vertex formalism should incorporate the aforementioned effects of the presence of an O7$^+$-plane: $\mathbb{Z}_2$ orbifold action and monodromy cut. To this end, we introduce two novel prescriptions for an O$7^+$-plane: \begin{itemize}
    \item[(i)] the contribution from a $\mathbb{Z}_2$ orbifold fixed point attached to a 5-brane, and 
    \item[(ii)] the contribution from the intersection between a 5-brane and the monodromy cut. 
\end{itemize}
We argue that the topological vertex formalism remains valid for 5-brane web diagrams containing an O7$^+$-plane, provided the contributions associated with these two types of ``vertices'' are assigned appropriately.

Before presenting our proposal, we motivate our claim by recalling the perspective that an O7$^+$-plane can be regarded as a $\mathbb{Z}_2$ orbifold with four frozen D7-branes:
\begin{align}
\text{``\,O7}^+ \,\sim\, \mathbb{Z}_2\, +\, 4 \,\text{D7\,.''}
\label{eq:O7=Z2+4D7}
\end{align}
This viewpoint was already implemented in the topological vertex formalism for 5-brane web diagrams with an O7$^+$-plane {\it not} attached to any 5-brane \cite{Kim:2024ufq}. Such a 5-brane configuration leads to 5d $\mathcal{N}=1$ SO($2N$) gauge theories with fundamental hypermultiplets. As shown in \cite{Kim:2024ufq}, the corresponding instanton partition function is computed from the topological vertex formalism based on a $\mathbb{Z}_2$ orbifold and four frozen D7-branes.

For the web where an O$7^+$-plane is attached an NS5-brane, we need a generalization for the $\mathbb{Z}_2$ orbifold and four D7-branes. 
In this paper, the contributions from these four frozen D7-branes are collectively consolidated and reformulated in terms of the intersection between a 5-brane and the monodromy cut. We subsequently demonstrate that our proposal reproduces the correct partition function for 5d $\mathcal{N}=1$ SO($2N$) gauge theory, thereby affirming its consistency with—and natural generalization of—the approach presented in~\cite{Kim:2024ufq}. 
In the following, we detail the new prescriptions for the two types of contributions mentioned above.

\subsubsection*{$\mathbb{Z}_2$-vertex}
Suppose an NS5-brane, or more generally, a $(p,1)$ 5-brane is attached to the $\mathbb{Z}_2$ orbifold fixed point, where the O7$^+$-plane is located. In this case, we propose that ``$\mathbb{Z}_2$-vertex,'' given in terms of the operator formalism of topological vertex as  
\begin{align}
\boxed{
V_{\mu}^{\mathbb{Z}_2} \equiv 
\left\langle 0 \left| \mathbb{O}_{\mathbb{Z}_2} \right| \mu \right\rangle,
\qquad 
\mathbb{O}_{\mathbb{Z}_2} \equiv \exp \left( - \sum_{n=1}^{\infty} \frac{1}{2n} (J_n)^2 \right)
,}
\label{eq:Z2-vertex}
\end{align}
should be assigned to the orbifold fixed point. The notation used here is explained in the following.

First, $\mu$ is the Young diagram assigned to the $(p,1)$ 5-brane that attaches to the orbifold fixed point, and  $\ket{\mu}$ is the state labeled by this Young diagram $\mu$ represented in the Frobenius basis. A set of such states forms an orthonormal basis, 
\begin{equation}
    \langle \mu | \nu \rangle = \delta_{\mu \nu} \ , 
    \label{eq:orthonormal}
\end{equation}
and satisfies the completeness relation,
\begin{equation}
    \sum_\mu\ \ket{\mu} \! \bra{\mu}  =\mathbbm{1}\ .
    \label{eq:comp}
\end{equation}

The operators $J_n$ ($n \in \mathbb{Z}$) in \eqref{eq:Z2-vertex} obeys the commutation relation
\begin{align}\label{eq:Commutation-J}
[J_n,J_m] = n\,  \delta_{n+m,0}\ .
\end{align}
By acting $J_n$ ($n \ge 0$) on a state $\ket{\mu}$, we obtain a linear combination of states that are labeled by Young diagrams whose number of boxes is given by $|\mu|-n$. When $|\mu|<n$, it vanishes. Especially, when $J_n$ 
acts on the state labeled by the empty Young diagram, it always vanishes
\begin{align}
J_n \ket{0} = 0 \quad (n \ge 0)\ .\label{eq:Jn|0>=0}
\end{align}

We note that the $\mathbb{Z}_2$-vertex \eqref{eq:Z2-vertex} vanishes if the number of boxes in the Young diagram $\mu$ is odd. Also, by generalizing the discussion in Appendix B in \cite{Hayashi:2020hhb}, we find that the $\mathbb{Z}_2$-vertex should be invariant under the simultaneous transformation,
\begin{align}\label{eq:ket-tr}
J_n \to - J_n, \qquad \ket{\mu} \to (-1)^{|\mu|} \ket{\mu^T}.
\end{align}
Thus, we find, from the definition \eqref{eq:Z2-vertex}, that the $\mathbb{Z}_2$-vertex satisfies the following identity: 
\begin{align}
 V_{\mu^T}^{\mathbb{Z}_2} = (-1)^{|\mu| }V_{\mu}^{\mathbb{Z}_2} = V_{\mu}^{\mathbb{Z}_2}.
 \label{eq:V-tr}
\end{align}

The proposal for the $\mathbb{Z}_2$-vertex \eqref{eq:Z2-vertex} is motivated by the ``O-vertex'' operator \cite{Hayashi:2020hhb, Nawata:2021dlk}
\begin{align}
\left\langle 0 \left| \exp \left( \sum_{n=1}^{\infty} - \frac{1}{2n} \frac{1+q^n}{1-q^n} J_{2n} + \frac{1}{2n} (J_n)^2 \right)\right| \mu \right\rangle,
\label{eq:O5-vertex}
\end{align}
which is proposed to be assigned to the intersection between a 5-brane and an O5-plane.
Going through the computation in Appendix C in \cite{Kim:2024ufq}, we find an implication that, schematically, the first term in \eqref{eq:O5-vertex} corresponds to ``4 {\it frozen} D7-branes'' in \eqref{eq:O7=Z2+4D7}, while the second term in \eqref{eq:O5-vertex} corresponds to ``$\mathbb{Z}_2$'' in \eqref{eq:O7=Z2+4D7}.
More precisely, the first term in the exponent in \eqref{eq:O5-vertex} is responsible for the contribution from the string connecting a color brane and one of the four frozen D5-branes, while the second term in the exponent in \eqref{eq:O5-vertex} is responsible for the contribution including the string connecting one color brane and its mirror image. Thus, it is natural to expect that the $\mathbb{Z}_2$-vertex is analogous to the second term in \eqref{eq:O5-vertex}. 

The significant difference would be that the contribution from the vector multiplet of SO($2N$) gauge group, which provides non-trivial factors in the denominators, is reproduced from the O-vertex, while we would like to reproduce the contribution from the hypermultiplet in symmetric tensor representation, which provides non-trivial factors in the numerators, from $\mathbb{Z}_2$-vertex. Because of this difference, we introduce a minus sign in the exponent in \eqref{eq:Z2-vertex} in contrast to \eqref{eq:O5-vertex}. 
We claim that this sign differentiates two distinct roles of the $\mathbb{Z}_2$ action on a 5-brane web: One is the rotation around an O7$^+$-plane, and the other is the reflection along an O5-plane.

\begin{figure}[ht]
\centering
\begin{tikzpicture}[
    dot/.style = {circle, draw, fill=black, inner sep=0pt, minimum size=2pt},
    smalldot/.style = {circle, draw, fill=black, inner sep=0pt, minimum size=1pt},
    label/.style = {text height=1.5ex, text depth=0.25ex},
    bigdot/.style = {circle, draw, fill=black, inner sep=0pt, minimum size=6pt},
]

\draw (0,0) circle (0.1cm);
\draw[-{Latex}] (0,0.1) -- (0,0.8);  
\draw (0,0.1) -- (0,1.5); 
\draw (0,1.5) -- (-0.5,2);
\draw (0,1.5) -- (1,1.5);
\node[left] at (0,0.8) {$\mu$};
\node[right] at (0,0.8) {Q};
\node[below] at (0,0) {$\mathbb{Z}_2$ vertex};
\node[below] at (0,-0.5) {(O7$^+$-plane)};

\begin{scope}[xshift=6cm]

\draw (0,0) circle (0.1cm);
\draw[-{Latex}] (0,0.1) -- (0,0.8);  
\draw (0,0.1) -- (0,1.5); 
\draw (0,1.5) -- (-1,1.5);
\draw (0,1.5) -- (0.5,2);

\node[left] at (0,0.8) {$\mu$};
\node[right] at (0,0.8) {Q};
\node[below] at (0,0) {$\mathbb{Z}_2$ vertex};
\node[below] at (0,-0.5) {(O7$^+$-\text{plane})};
\end{scope}
\end{tikzpicture}
\caption{
The powers $n$ for the framing factor assigned to the edges connected to the $\mathbb{Z}_2$ orbifold fixed point (O7$^+$-plane): 
Left: $n=0$. Right: $n=1$.
}
\label{fig:O7-framing}
\end{figure}
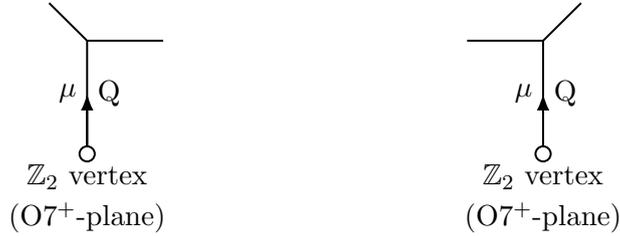

In addition to the assignment to the orbifold fixed point, we also propose the assignment to the edge connecting the orbifold fixed point and a trivalent vertex. Denoting the Young diagram assigned to this edge as $\mu$ and the corresponding K\"ahler parameter as $Q$, we expect the corresponding contribution to be of the form, 
\begin{align}
(-Q)^{|\mu|} f_{\mu}{}^n ,
\label{eq:framing-O7}
\end{align}
analogous to the contributions from other edges connecting two trivalent vertices.
Here, the framing factor $f_\mu$ takes the form  
\begin{align}
    f_\mu:=(-1)^{|\mu|}q^{\frac12 \kappa(\mu)},\qquad \kappa(\mu):=2\sum_{(i,j)\in\mu}(j-i)\ . \label{framing}
\end{align}
We note that the factor $(-1)^{|\mu|}$ included in \eqref{eq:framing-O7} is negligible due to  the property \eqref{eq:V-tr}. Thus, this can also be replaced by $Q^{|\mu|} q^{\frac12 n \kappa(\mu)}$. We propose that the power $n$ in the framing factor \eqref{eq:framing-O7} is determined by treating the orbifold fixed point as an effective trivalent vertex, created by adding a D5-brane extending leftward from it. 
Especially, when the considered edge is connected to a color D5-brane, $n$ equals $0$ or $1$ depending on the attached direction of the color brane, as illustrated in Figure \ref{fig:O7-framing}. Although this rule for $n$ lacks an a priori justification, it produces the expected partition function.

\subsubsection*{FD-vertex}
\begin{figure}[ht]
\centering
\begin{tikzpicture}[
    dot/.style = {circle, fill=black, inner sep=0pt, minimum size=2pt},
    hollow/.style = {circle, draw=black, fill=white, inner sep=0pt, minimum size=2pt},
    arrow/.style = {-{Latex}},
    bigdot/.style = {circle, fill=black, inner sep=0pt, minimum size=6pt}
]
\draw[dashed] (-3,0) -- (2.9,0);
\draw (3,0.1) -- (3,1.5);
\draw (3,0) circle (0.1cm);
\node[anchor=west] at (3.2,0) {O7$^{+}$\text{-plane}};
\draw (-2.5,-1.5) -- (0,0) -- (2.5,1);
\draw[arrow] (0,0) -- (1.5,0.6);
\draw[arrow] (-2.5,-1.5) -- (-1,-0.6);
\node[above left] at (-1.2,-0.7) {$\mu$};
\node[above right] at (1.2,0.7) {$\nu$};
\node at (-0.2,0.3) {$V_{\mu\nu}^{FD}$};
\draw [fill] (0,0) circle (0.1cm);
\end{tikzpicture}
\caption{
The FD-vertex and Young diagram assignment.}
\label{fig:FD-vertex}
\end{figure}
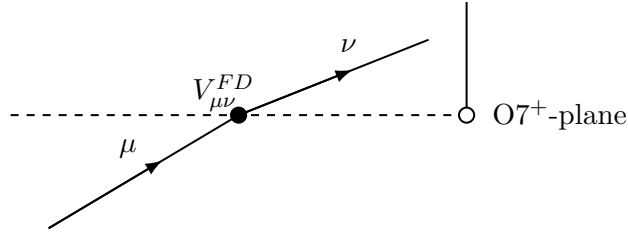

Now we consider the intersection between a 5-brane and the monodromy cut. We propose to associate the following ``FD-vertex''--short for ``Frozen D-brane vertex''--to the intersection of a 5-brane with the monodromy cut induced by the O7$^+$-plane:
\begin{align}
\boxed{
V^{\mathrm{FD}}_{\mu\nu} \equiv q^{\frac{\kappa(\mu)}{2}}
\bra{\mu}
\mathbb{O}_{\mathrm{FD}}^- \mathbb{O}_{\mathrm{FD}}^+
\ket{\nu},
\quad 
\mathbb{O}_{\mathrm{FD}}^{\pm} \equiv 
\exp \left( -\sum_{n=1}^{\infty} \frac{1}{2n} \frac{1+q^n}{1-q^n} J_{\pm 2n} \right)
}
\label{eq:FD-vertex}
\end{align}
Here, $\mu$ and $\nu$ are Young diagrams assigned to the FD-vertex in clockwise order from the point of the O7$^+$-plane, as depicted in Figure \ref{fig:FD-vertex}. 

This proposal is again motivated by the O-vertex \eqref{eq:O5-vertex}. Since the first term in \eqref{eq:O5-vertex} corresponds to four frozen D-branes as discussed before, it is natural to expect that the FD-vertex is analogous to this term. Indeed, if $\mu=\varnothing$, \eqref{eq:FD-vertex} reduces to 
\begin{align}\label{eq:red-FD}
V^{\mathrm{FD}}_{\varnothing \nu}
=
\bra{0}
\mathbb{O}_{\mathrm{FD}}^{+}
\ket{\nu} ,
\end{align}
which is identical to the contribution from the frozen D-branes in the O-vertex.
Similar to the property of $V_{\mu \nu}^{\mathbb{Z}_2}$
, one finds
\begin{align}
V_{\mu^T \nu^T}^{\mathrm{FD}} 
= q^{-\kappa(\mu)} (-1)^{|\mu|+|\nu|}V_{\mu \nu}^{\mathrm{FD}}
= q^{-\kappa(\mu)} V_{\mu \nu}^{\mathrm{FD}}.
\end{align}

We further propose an assignment for the edge connecting the FD-vertex and a trivalent vertex. As usual, we expect it to take the form given in~\eqref{eq:framing-O7}, identical to an edge connecting two trivalent vertices. Analogous to the previous case, we postulate the power $n$ of the framing factor is determined by treating the FD-vertex as an effective trivalent vertex, realized by adding a D5-brane extending from the FD-vertex to the left.

\subsection{5d \texorpdfstring{$\mathcal{N}=1$ SU$(N)_k$}{N=1 SU(N)} gauge theory with a symmetric hypermultiplet}

\begin{figure}[ht]
\centering
\begin{tikzpicture}[
    line/.style = {thick},
    dot/.style = {circle, fill=black, inner sep=0pt, minimum size=2pt},
    label/.style = {font=\small}
]

\draw[line] (-2,-3) -- (-2,3); 
\draw[line] (0,0.1) -- (0,3);   
\draw[line] (0,-3) -- (0,-0.1);
\draw[line] (2,-3) -- (2,3);   

\node at (-2.5, 2.5) {$N$};
\draw[line] (-2,2.6) -- (0,2.6);   
\draw[line] (-2,2.3) -- (0,2.3);   
\draw[line] (-2,2) -- (0,2); 

\draw[line] (0,0) circle (0.1cm);
\node at (0.5,0.5) {O7$^+$};

\draw[line] (0,-2) -- (2,-2);   
\draw[line] (0,-2.3) -- (2,-2.3);   
\draw[line] (0,-2.6) -- (2,-2.6); 
\end{tikzpicture}
\caption{The 5-brane web with O7$^+$-plane on which 5d SU($N$) gauge theory with a hypermultiplet in symmetric tensor representation is realized.}
\label{Fig:SU+Sym}
\end{figure}
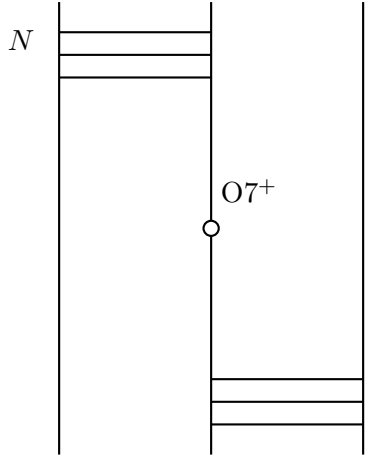

In order to demonstrate how our proposal works, we consider 5d $\mathcal{N}=1$ SU($N$) gauge theory at Chern-Simons level $k$ with a hypermultiplet in symmetric tensor representation as an example. The corresponding 5-brane web is depicted in Figure \ref{Fig:SU+Sym}. In this figure, all $(p,1)$ 5-branes are written vertically for simplicity. Also, note that this figure includes the mirror image.

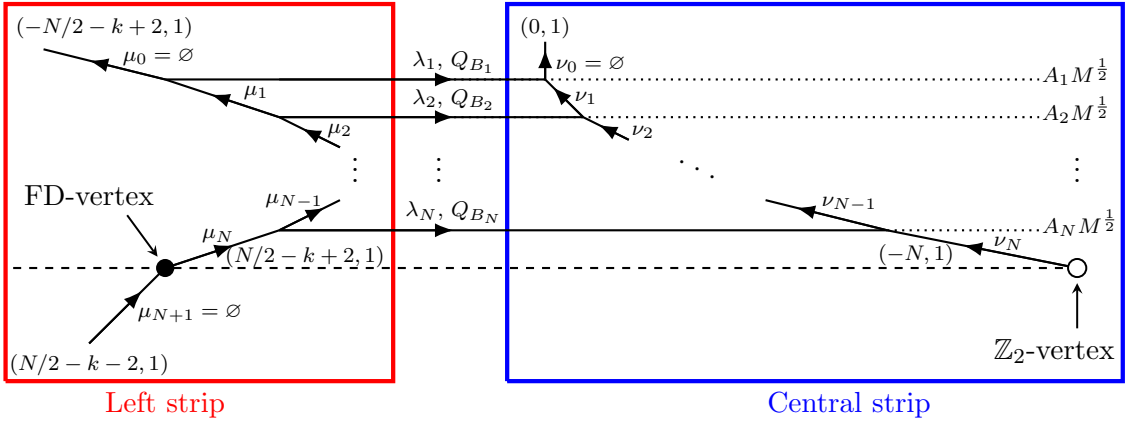
\begin{figure}[ht]
\centering
\begin{tikzpicture}[
    line/.style = {thick},
    dot/.style = {circle, fill=black, inner sep=0pt, minimum size=2pt},
    label/.style = {font=\small}
]
\draw[line,dashed] (-1,1) -- (13,1); 
\draw[line] (0,0) -- (1,1) -- (2.5,1.5) -- (3.5-0.2,2-0.1); 
\draw (3.5,2) node[above]{$\vdots$};
\draw[line] (3.5-0.2,2.5+0.1) -- (2.5,3) -- (1,3.5) -- (-1+0.4,4-0.1); 
\draw[-{Latex[length=2.5mm]}]  (0,0)--(0.7,0.7);
\draw[-{Latex[length=2.5mm]}]  (1,1)--(1.9,1.3);
\draw[-{Latex[length=2.5mm]}]  (2.5,1.5)--(3.5-0.4,2-0.2);
\draw[-{Latex[length=2.5mm]}]  (3.5-0.2,2.5+0.1)--(2.5+0.3,3-0.15);
\draw[-{Latex[length=2.5mm]}]  (2.5,3)--(1+0.6,3.5-0.2);
\draw[-{Latex[length=2.5mm]}]  (1,3.5)--(-1+1,4-0.25);
\draw (0.5,0.4) node[right]{\scriptsize $\mu_{N+1}=\varnothing$};
\draw (1.7,1.2) node[above]{\scriptsize $\mu_{N}$};
\draw (3.6-0.9,1.4+0.25) node[above]{\scriptsize $\mu_{N-1}$};
\draw (3.3,3+0.05) node[below]{\scriptsize $\mu_{2}$};
\draw (2.2,3.5+0.05) node[below]{\scriptsize $\mu_{1}$};
\draw (0.3,3.8) node[right]{\scriptsize $\mu_{0} = \varnothing$};
\draw (0,0) node[below]{\scriptsize $(N/2-k-2,1)$};
\draw (1.65,1.15) node[right]{\scriptsize $(N/2-k+2,1)$};
\draw (0+0.2,4-0.1) node[above]{\scriptsize $(-N/2-k+2,1)$};
\fill(1,1) circle [radius=1.2mm];
\draw (0,1.6+0.1) node[above]{FD-vertex};
\draw[->,>=stealth] (0.5,1.6+0.1)--(0.9,1.15);
\draw[line,ultra thick,red] (-1.6+0.5,-0.5) -- (-1.6+0.5,4.5) -- (4,4.5) -- (4,-0.5) -- (-1.6+0.5,-0.5);
\draw[red] (1,-0.5) node[below]{Left strip};
\draw[line] (13,1) -- (10.5,1.5) -- (8.5+0.4,2-0.1); 
\draw (8-0.2,2.3+0.1) node{$\cdot$};
\draw (8,2.3) node{$\cdot$};
\draw (8+0.2,2.3-0.1) node{$\cdot$};
\draw[line] (7.5-0.4,2.5+0.2)-- (6.5,3) -- (6,3.5) -- (6,4) ; 
\draw[-{Latex[length=2.5mm]}]  (13,1)--(11.5,1.3);
\draw[-{Latex[length=2.5mm]}]  (10.5,1.5)--(8.5+0.8,2-0.2);
\draw[-{Latex[length=2.5mm]}]  (7.5-0.4,2.5+0.2) -- (6.5+0.2,3-0.1);
\draw[-{Latex[length=2.5mm]}]  (6.5,3) -- (6+0.1,3.5-0.1);
\draw[-{Latex[length=2.5mm]}]  (6,3.5) -- (6,4-0.1);
\draw (12+0.1,1.55) node[below]{\scriptsize $\nu_{N}$};
\draw (9.5,1.5+0.35) node[right]{\scriptsize $\nu_{N-1}$};
\draw (6.5+0.5,3-0.2) node[right]{\scriptsize $\nu_{2}$};
\draw (6+0.25,3.5-0.25) node[right]{\scriptsize $\nu_{1}$};
\draw (6,4-0.3) node[right]{\scriptsize $\nu_{0}=\varnothing$};
\draw (11-0.12,0.9-0.01) node[above]{\scriptsize $(-N,1)$};
\draw (6,4-0.1) node[above]{\scriptsize $(0,1)$};
\filldraw [fill=white,draw=black](13,1) circle [radius=1.2mm];
\draw (13-0.3,0.2) node[below]{$\mathbb{Z}_2$-vertex};
\draw[->,>=stealth] (13,0.2)--(13,0.8);
\draw[line,ultra thick,blue] (5.5,-0.5) -- (5.5,4.5) -- (14.2-0.6,4.5) -- (14.2-0.6,-0.5) -- (5.5,-0.5);
\draw[blue] (10,-0.5) node[below]{Central strip};
\draw[line] (2.5,1.5) -- (10.5,1.5);
\draw (5-0.4,2) node[above]{$\vdots$};
\draw[line] (2.5,3) -- (6.5,3);
\draw[line] (1,3.5) -- (6,3.5);
\draw[-{Latex[length=2.5mm]}] (2.5,1.5) -- (5-0.2,1.5);
\draw[-{Latex[length=2.5mm]}] (2.5,3) -- (5-0.2,3);
\draw[-{Latex[length=2.5mm]}] (2.5,3.5) -- (5-0.2,3.5);
\draw (5-0.2,1.45) node[above]{\scriptsize $\lambda_{N}, \, Q_{B_N}$};
\draw (5-0.2,3-0.05) node[above]{\scriptsize $\lambda_{2}, \, Q_{B_2}$};
\draw (5-0.2,3.5-0.05) node[above]{\scriptsize $\lambda_{1}, \, Q_{B_1}$};
\draw[line,dotted] (10.5,1.5) -- (13-0.5,1.5);
\draw (13-0.6,1.5+0.1) node[right]{\scriptsize $A_NM^{\frac12}$};
\draw (13,2) node[above]{$\vdots$};
\draw[line,dotted] (5-0.2,3) -- (13-0.5,3);
\draw (13-0.6,3+0.1) node[right]{\scriptsize $A_2M^{\frac12}$};
\draw[line,dotted] (5-0.2,3.5) -- (13-0.5,3.5);
\draw (13-0.6,3.5+0.1) node[right]{\scriptsize $A_1M^{\frac12}$};
\end{tikzpicture}
\caption{Assignment of Young diagrams and decomposition of the diagrams into strips for the 5d SU($N$)$_k$ + 1 {\bf Sym}.}
\label{Fig:Z2+FDvertex}
\end{figure}

When we compute it, we focus on the fundamental region, which we choose to be the upper left half of the 5-brane web in Figure \ref{Fig:SU+Sym} in this paper. Then, we decompose it into the left strip and the central strip, which are connected by the color branes, as depicted in Figure \ref{Fig:Z2+FDvertex}.

Note that the K\"ahler parameters are given in terms of the gauge theory parameters as  
\begin{align}\label{eq:Kahler}
A_s = e^{-a_s}, 
\quad
M = e^{-m}
\quad
Q_{Bs} = (-1)^{N-1} Q_I M^{-\frac12 N-1} A_s^{\frac12 N + k - 2s} \prod_{r=1}^{s-1} A_t^2,
\end{align}
where $a_s$ ($s=1, \cdots N$) are Coulomb moduli, which satisfy the traceless condition $\sum_s a_s=0$, $m$ is the mass for the symmetric tensor, and $Q_I$ is the instanton factor.

The topological string partition function is given by gluing the amplitudes for the left strip and the central strip in the form
\begin{align}\label{eq:Z-top}
Z_{\text{top}}
&= \sum_{\mathbf{\lambda}} \prod_{s=1}^N (-Q_{B s}) ^{|\lambda_s|} f_{\lambda_s}^{-\frac{N}{2} - k + 1  + 2 s}
Z_{\text{left}} Z_{\text{central}} \ .
\end{align}
Here, by using our proposal, the amplitude for the central strip and for the left strip are respectively given by
\begin{align}
Z_{\text{central}}\label{eq:def-left-central}
&:= \sum_{ \vec{\mathbf{\nu}} } 
\left( 
\prod_{r=1}^{N-1} (-A_{r} A_{r+1}^{-1})^{|\nu_r|} 
f_{\nu_r}\right) 
(A_N M^{\frac12})^{|\nu_N|}
q^{\frac12 \kappa(\nu_N)}
V^{\mathbb{Z}_2}_{\nu_N} \prod_{s=1}^N C_{\nu_{s-1} \nu_s^T \lambda_s^T},
\cr
Z_{\text{left}} 
&:= \sum_{ \vec{\mathbf{\mu}} } 
\left( 
\prod_{r=1}^{N-1} (-A_{r} A_{r+1}^{-1})^{|\mu_r|} f_{\mu_r}^{-1}
\right) (A_N M^{\frac12})^{|\mu_N|}
V^{\mathrm{FD}}_{\varnothing \mu_N} \prod_{s=1}^N C_{\mu_{s}^T \mu_{s-1} \lambda_s}, \
\end{align}
where we define $\vec{\nu} = (\nu_1, \cdots, \nu_N)$, $\vec{\mu} = (\mu_1, \cdots, \mu_N)$, and $\nu_0 = \mu_0 = \varnothing$.
Here, the framing factor $f_\lambda$ is defined in \eqref{framing} and the vertex factor $C_{\mu\nu\lambda}$ can be expressed with (skew) Schur functions as 
\begin{align}
\begin{tikzpicture}[scale=0.5]
\draw (0,0) -- (-1,-1) ;
\draw[->,>=Latex] (-0.7,-0.7) -- (-0.71,-0.71) node [above left] {$\mu$};
\draw (0,0) -- (0,1.4) ;
\draw[->,>=Latex] (0,0.9) -- (0,0.91) node [above left] {$\nu$};
\draw (0,0) -- (1.4,0) ;
\draw[->,>=Latex] (0.9,0) -- (0.91,0) node [above right] {$\lambda$};
\node at (2.5,0) [right]  {$\mathbf{\Leftrightarrow}$} ;
\node at (4,0) [right]  {$C_{\mu\nu\lambda} =q^{\frac{1}{2}\kappa(\nu)}
s_{\lambda^T}(q^{-\rho})\sum_{\sigma}s_{\mu^T/\sigma}(q^{-\rho-\lambda})s_{\nu/\sigma}(q^{-\rho-\lambda^T})\ ,$};
\end{tikzpicture}
\end{align}
where the Young diagrams $\mu, \nu, \lambda$ are arranged in clockwise order, and all arrows from this vertex are taken to be outgoing. If an arrow is incoming, the corresponding Young diagram should be replaced by its transpose.

Substituting these definitions above into \eqref{eq:def-left-central} and by absorbing the dependence on the K\"ahler parameters $A_r$, $M$ into the skew Schur functions by using the identity
\begin{align}
Q^{|\lambda|-|\mu|}  s_{\lambda/\mu}(\vec{x}) = s_{\lambda/\mu}(Q \vec{x})\ , 
\end{align}
we find the two strip amplitudes above are rewritten, respectively, as
\begin{align}
Z_{\text{central}} 
&= \prod_{s=1}^N s_{\lambda_s}(q^{-\rho}) 
\times
\sum_{\vec{\nu}, \vec{\sigma}} V_{\nu_N}^{\mathbb{Z}_2}
\prod_{s=1}^N s_{\nu_{s-1}^T/\sigma_s}(A_s^{-1} M^{-\frac12}) 
s_{\nu_{s}^T/\sigma_s}(A_s M^{\frac12}) ,
\cr
Z_{\text{left}} 
&= \prod_{s=1}^N s_{\lambda_s^T}(q^{-\rho}) 
\times 
\sum_{\vec{\mu}, \vec{\sigma}} V_{\varnothing \mu_N}^{\text{FD}}
\prod_{s=1}^N s_{\mu_{s-1}/\sigma_s} (A_s^{-1} M^{-\frac12}) 
s_{\mu_{s}/\sigma_s}(A_s M^{\frac12}) .
\end{align}

One can evaluate these strip amplitudes by applying sequently the Cauchy identity
\begin{align}
& \sum_\lambda s_{\lambda/\sigma}( Q q^{-\rho-\mu})\,s_{\lambda/\tau}( Q' q^{-\rho-\nu})
= R^{-1}_{\mu \nu}(Q Q') \!\sum_{\lambda} \,s_{\tau/\lambda}(Q q^{-\rho-\mu})\,s_{\sigma/\lambda}(Q' q^{-\rho-\nu})\ . \label{Schur-id-spec}
\end{align}
Following the computation developed in various works computing strip amplitudes, including \cite{Iqbal:2007ii}, we find
\begin{align}
Z_{\text{central}}
=   
\prod_{s=1}^N s_{\lambda_s}(q^{-\rho}) 
\times 
\prod_{1\le s<t \le N} R^{-1}_{\lambda_{s} \lambda_{t}^T}(A_s A_t^{-1}) 
\times \tilde{Z}^{\mathbb{Z}_2} \ ,
\cr
Z_{\text{left}}
=    
\prod_{s=1}^N s_{\lambda_s^T}(q^{-\rho}) 
\times 
\prod_{1\le s<t \le N} R^{-1}_{\lambda_{s} \lambda_{t}^T}(A_s A_t^{-1})
\times
\tilde{Z}^{\text{FD}}  \ ,
\label{eq:LC-intermed}
\end{align}
with
\begin{align}
\tilde{Z}^{\mathbb{Z}_2} 
:= & 
\sum_{\vec{\nu}} V^{\mathbb{Z}_2}_{\nu_{N}} \prod_{s=1}^{N} \,s_{\nu_s/\nu_{s-1}}( A_s M^{\frac12} q^{-\rho-\lambda_{s}} ) , 
\cr
\tilde{Z}^{\text{FD}} 
:= &
\sum_{\vec{\mu}} V^{\text{FD}}_{\varnothing \mu_{N}} \prod_{s=1}^{N} \,s_{\mu_s^T/\mu_{s-1}^T}( A_s M^{\frac12} q^{-\rho-\lambda_{s}} ) \ .
\label{eq:def-ZZ2-ZFD}
\end{align}
Here, we have extracted the factors including the contributions from $\mathbb{Z}_2$ vertex and the FD vertex, respectively, as defined in \eqref{eq:def-ZZ2-ZFD}. The remaining factors in \eqref{eq:LC-intermed} agree with the strip amplitudes that are obtained by replacing these vertices by a 5-brane that extends to infinity.

In the following, we focus on the factors \eqref{eq:def-ZZ2-ZFD} and rewrite them in the operator formalism. We note that the skew Schur functions can be rewritten as an operator expression, 
\begin{align}
 s_{\lambda/\mu}(\vec{x})=\bra{\mu}\Gamma_+(\vec{x})\ket{\lambda}=\bra{\lambda}\Gamma_-(\vec{x})\ket{\mu}  \ ,  
\end{align} 
with
\begin{equation}\label{eq:Gamma-pm}
    \Gamma_\pm (\vec{x}):=\exp\lt(\sum_{n=1}^\infty \frac{1}{n}\sum_i x_i^n J_{\pm n}\rt)~. 
\end{equation}
It follows from the completeness relation of the basis \eqref{eq:comp} that these factors can be given in a compact form as 
\begin{align}\label{Z-CL-operator}
\tilde{Z}^{\mathbb{Z}_2}
&= \bra{0}\mathbb{O}_{\mathbb{Z}_2}\prod_{s=1}^N\Gamma_-(A_s M^{\frac12} q^{-\rho-\lambda_{s}})\ket{0}\ ,
\cr
\tilde{Z}^{\text{FD}}
&= \bra{0}\mathbb{O}^+_{\text{FD}}\prod_{s=1}^N\Gamma_-(A_s M^{\frac12} q^{-\rho-\lambda_{s}})\ket{0}\ .
\end{align}

These operator expectation values can be computed by using the variation of the Baker-Campbell-Hausdorff formula
\begin{align}
e^X e^Y = e^Y e^{X+[X\,,\,Y]+\frac12 \lt[[X\,,\,Y]\,,\,Y\rt] + \frac{1}{3!} \lt[\lt[[X\,,\,Y]\,,\,Y\rt],\,Y\rt] + \cdots}\ 
\end{align}
with the identification
\begin{align}
e^X = \mathbb{O}_{\mathbb{Z}_2} \text{ or }\mathbb{O}_{\text{FD}}^+,
\qquad
e^Y = \prod_{s=1}^N\Gamma_-(A_s M^{\frac12} q^{-\rho-\lambda_{s}}).
\end{align}
Essentially the same computations have already been done in Appendix C of \cite{Kim:2024ufq}. Following the same process, we find 
\begin{align}\label{eqZ-FD-result}
\tilde{Z}^{\mathbb{Z}_2}
& = 
\prod_{s,t=1}^N 
R_{  \lambda_{s} \lambda_{t} }^{\frac12} (A_s A_t M) \ ,
\cr
\tilde{Z}^{\text{FD}}
&= 
\prod_{f=1}^4 \prod_{s=1}^N \prod_{\ell= \pm 1} R^{\frac12}_{\lambda_s \varnothing} (A_s M^{\frac12} \mathsf{M}_f^{\ell})\ ,
\end{align}
where we have introduced 
\begin{align}\label{eq:def-Mf}
\mathsf{M}_f := e^{-\mathsf{m}_f} = \{~ \pm 1\ , ~ \pm q^{\frac12} ~\}\ ,
\end{align}
with $\mathsf{m}_f$ being defined in \eqref{eq:tuned-mass}.

Combining all the results given in \eqref{eq:Z-top}, \eqref{eq:LC-intermed}, and \eqref{eqZ-FD-result}, we finally obtain the topological string amplitude as
\begin{align}\label{eq:Ztop-final}
Z_{\text{top}}
& = \sum_{\mathbf{\lambda}} \prod_{s=1}^N (-Q_{B s}) ^{|\lambda_s|} f_{\lambda_s}^{-\frac{N}{2} - k + 1  + 2 s} \cdot
\prod_{s=1}^N s_{\lambda_s}(q^{-\rho}) s_{\lambda_s^T}(q^{-\rho})  \cdot
\prod_{s<t} R^{-2}_{\lambda_s \lambda_t^T} (A_s A_t^{-1}) 
\cr
& \qquad \times 
\prod_{s,t=1}^N 
R_{  \lambda_{s} \lambda_{t} }^{\frac12} (A_s A_t M)
\cdot
\prod_{f=1}^4 \prod_{s=1}^N \prod_{\ell = \pm 1}
R^{\frac12}_{\lambda_s \varnothing} (A_s M^{\frac12} \mathsf{M}_f^{\ell})\ .
\end{align}

In the following, we discuss the validity of this expression \eqref{eq:Ztop-final}.
First, we check that it correctly reproduces the expected perturbative part, 
\begin{align}\label{eq:expert}
Z_{\mathrm{pert}} := 
\text{PE} \left[ \frac{q}{(1-q)^2} \left( 2 \sum_{s<t} A_s A_t^{-1} - \sum_{s \le t} M A_s A_t \right) \right].
\end{align}
This perturbative part should be reproduced in the weak coupling limit $Q_I \to 0$ of \eqref{eq:Ztop-final}. 
In this limit, since $Q_{Bs} \to 0$ due to the parametrization \eqref{eq:Kahler}, only the term with $\lambda_s = \varnothing$ remains non-zero to give
\begin{align}\label{eq:qto0limit}
\lim_{Q_I \to 0} Z_{\text{top}} =
& 
\prod_{s<t} R^{-2}_{\varnothing\varnothing} (A_s A_t^{-1}) 
\times 
\prod_{s,t=1}^N 
R_{\varnothing\varnothing}^{\frac12} (A_s A_t M)
\times 
\prod_{f=1}^4 \prod_{s=1}^N \prod_{\ell = \pm 1}
R^{\frac12}_{\varnothing\varnothing} (A_s M^{\frac12} \mathsf{M}_f^{\ell}).
\end{align}
Using the identity 
\begin{align}\label{eq:idRnn}
R_{\varnothing\varnothing}(X) 
= \mathrm{exp} \left( - \sum_{n=1}^{\infty} \frac1n \frac{q^n}{(1-q^n)^2} X^n \right),
\end{align}
the third factor in the expression \eqref{eq:qto0limit} gives
\begin{align}\label{eq:third-exp}
& \prod_{f=1}^4 \prod_{s=1}^N \prod_{\ell = \pm 1}
R^{\frac12}_{\varnothing\varnothing} (A_s M^{\frac12} \mathsf{M}_f^{\ell})
= \exp \left( - \frac12 \sum_{s=1}^N \sum_{n=1}^{\infty} \frac1n \frac{q^n M^{\frac{n}{2}} A_s^n}{(1-q^n)^2}   \sum_{f=1}^4 \sum_{\ell = \pm 1} \mathsf{M}_f^{\ell n} \right).
\end{align}
Using the explicit expressions \eqref{eq:def-Mf} for $\mathsf{M}_f$, the summation over the indices $f$ and $\ell$ can be done explicitly to give 
\begin{align}
\sum_{f=1}^4 \sum_{\ell = \pm 1} \mathsf{M}_f^{\ell n} 
= \left( 1 + (-1)^n \right) \left( 2 + q^{\frac{n}{2}} + q^{-\frac{n}{2}} \right)
= \left\{
\begin{array}{ll}
2 q^{-k} (1 + q^k )^2 & \text{for } n = 2k \\
0 & \text{for } n = 2k+1.
\end{array}
\right.
\end{align}
This indicates that only the terms with even $n$ in \eqref{eq:third-exp} remain non-zero to give
\begin{align}
\prod_{f=1}^4 \prod_{s=1}^N \prod_{\ell = \pm 1}
R^{\frac12}_{\varnothing\varnothing} (A_s M^{\frac12} \mathsf{M}_f^{\ell})
=& \exp \left( - \frac12 \sum_{s=1}^N \sum_{k=1}^{\infty} \frac{1}{k}   \frac{q^{k}M^{k} A_s^{2k}}{(1-q^{2k})^2} 
(1 + q^{k})^2 
\right)
\cr
= & \mathrm{PE} \left[ - \frac{q}{2(1-q)^2} \sum_{s=1}^N M A_s^2
\right].
\end{align}
Combined this with the first and the second factor in \eqref{eq:qto0limit}, which are given, respectively, as
\begin{align}
\prod_{s<t} R^{-2}_{\varnothing\varnothing} (A_s A_t^{-1}) 
&= \text{PE} \left[ \frac{2 q}{(1-q)^2} \sum_{s<t} A_s A_t^{-1} \right] ,
\cr
\prod_{s,t=1}^N 
R_{\varnothing\varnothing}^{\frac12} (A_s A_t M)
&= \text{PE} \left[ - \frac{q}{(1-q)^2} \sum_{s \le t}^N M A_s A_t 
+ \frac{q}{2(1-q)^2} \sum_{s=1}^N M A_s^2
\right],
\end{align}
it is straightforward to check that \eqref{eq:qto0limit} agrees with the expected perturbative part \eqref{eq:expert}. 

Next, we check that the expression \eqref{eq:Ztop-final} also reproduces the expected instanton partition function, which we have given in \eqref{eq:Nek-SU+Sym-final}.
The instanton part should be obtained by dividing the expression \eqref{eq:Ztop-final} by the perturbative part, which we have confirmed to agree with \eqref{eq:qto0limit}. That is,
\begin{align}\label{eq:Z-inst}
Z_{\text{inst}}
&= \frac{Z_{\text{top}}}{\lim\limits_{Q_I \to 0} Z_{\text{top}}}
\cr
&= \sum_{\mathbf{\lambda}} \prod_{s=1}^N \left( (-1)^{N-1} Q_I M^{-\frac{N}{2}-1} A_s^{\frac{N}{2} + k - 2s} \prod_{r=1}^{s-1} A_r^2 \right) ^{|\lambda_s|} q^{- \frac14(N + 2k - 2 - 4 s) \kappa(\lambda_s)} 
\cr
& \qquad \times
\prod_{s=1}^N N^{-1}_{\lambda_s \lambda_s}(1) 
\times
\prod_{1 \le s<t \le N} N^{-2}_{\lambda_t \lambda_s} (A_s A_t^{-1}) 
\cr
& \qquad \times 
\prod_{s,t=1}^N 
N_{  \lambda_{s}^T \lambda_{t} }^{\frac12} (A_s A_t M)
\times
\prod_{f=1}^4 \prod_{s=1}^N \prod_{\ell = \pm 1}
N^{\frac12}_{\lambda_s^T \varnothing} (A_s M^{\frac12} \mathsf{M}_f^{\ell}).
\end{align}
Here, we have used the identities,  
\begin{align}
&R_{\mu \nu} (X) 
= R_{\varnothing\varnothing} (X)  N_{\mu^T \nu} (X)\ ,
\cr
&s_{\lambda_s}(q^{-\rho}) s_{\lambda_s^T}(q^{-\rho})
= (-1)^{|\lambda_s|} N^{-1}_{\lambda_s \lambda_s}(1)\ ,
\end{align}
as well as the parameterization \eqref{eq:Kahler} for $Q_{Bs}$ and the definition of the framing factor in \eqref{framing}.

Rewriting the Nekrasov factors in the expression \eqref{eq:Z-inst} using the relation
\begin{align}\label{eq:def-nekt-1}
N_{\lambda\nu}(X)
&= 
X^{\frac12 \big(|\lambda|+|\nu|\big)} q^{\frac14 \big(\kappa(\lambda) - \kappa(\nu)\big)}
\tilde{n}_{\lambda\nu}(x) ,
\qquad
X:=e^{-x}\ , 
\end{align}
we have
\begin{align}
&\prod_{s=1}^N N^{-1}_{\lambda_s \lambda_s}(1) 
= \prod_{s=1}^N \tilde{n}^{-1}_{\lambda_s \lambda_s}(0) \ ,
\cr
&\prod_{s<t} N^{-2}_{\lambda_t \lambda_s} (A_s A_t^{-1}) 
= 
\prod_{s=1}^N q^{\frac12 (N-2s+1) \kappa(\lambda_s)}
\left( A_s^{-N+2s} \prod_{t=1}^{s} A_t^{-2} \right)^{|\lambda_s|}
\cdot
\prod_{s<t} \tilde{n}^{-2}_{\lambda_t \lambda_s} (a_s - a_t ) \ 
 \cr
& \qquad\qquad\qquad
= 
\prod_{s=1}^N q^{\frac12 (N-2s+1) \kappa(\lambda_s)}
\left( (-1)^{N-1} A_s^{-N+2s} \prod_{t=1}^{s} A_t^{-2} \right)^{|\lambda_s|}
\cdot
\prod_{s \neq t} \tilde{n}^{-1}_{\lambda_t \lambda_s} (a_s - a_t ) \ , 
\cr
&\prod_{s,t=1}^N 
N_{ \lambda_{s}^T \lambda_{t} }^{\frac12} (A_s A_t M)
 = \prod_{s=1}^N q^{-\frac{N}{4} \kappa(\lambda_s)}
 ( M A_s) ^{\frac{N}{2} |\lambda_s|}
\cdot 
\prod_{s,t=1}^N 
\tilde{n}_{\lambda_{s}^T \lambda_{t} }^{\frac12} (a_s+a_t+m)\ , 
\cr
&\prod_{f=1}^4 \prod_{s=1}^N \prod_{\ell = \pm 1}
N^{\frac12}_{\lambda_s^T \varnothing} (A_s M^{\frac12} \mathsf{M}_f^{\ell})
 = \prod_{s=1}^N 
 q^{-\kappa(\lambda_s)}
 (M A_s^2)^{|\lambda_s|} 
 \prod_{s=1}^N \prod_{f=1}^{4}  \tilde{n}^{\frac12}(a_s + \frac12 m \pm  \mathsf{m}_f)\ .
\end{align}
These lead to
\begin{align}\label{eq:Z-inst2}
Z_{\text{inst}}
&= \sum_{\mathbf{\lambda}} \prod_{s=1}^N Q_I^{|\lambda_s|} 
\left( A_s^{|\lambda_s|} q^{-\frac12 \kappa(\lambda_s)} \right) ^k
\cdot
\prod_{s,t=1}^N 
\tilde{n}^{-1}_{\lambda_t \lambda_s} (a_s-a_t)\,
\tilde{n}_{  \lambda_{s}^T \lambda_{t} }^{\frac12} (a_s+a_t+m)
\cr
& \quad \times
\prod_{f=1}^4 \prod_{s=1}^N 
\tilde{n}^{\frac12}_{\lambda_s^T \varnothing} (a_s + \frac12 m \pm \mathsf{m}_f)\ ,
\end{align}s
which agrees with \eqref{eq:Nek-SU+Sym-final}, hence giving support for our proposal in section \ref{sec:proposal}.

\subsection{Further consistency checks}\label{subsec:fcon}

In the previous subsection, we confirmed that Nekrasov's instanton partition function for 5d $\mathcal{N}=1$ SU($N$) gauge theory with a hypermultiplet in the symmetric tensor representation is reproduced by applying our proposal to the 5-brane web diagram given in Figure \ref{Fig:Z2+FDvertex}, which includes the O7$^+$-plane attached to an NS5-brane. This 5-brane web diagram corresponds to the phase, 
\begin{align}
a_1 + \frac12m > a_2 + \frac12m> \cdots > a_N + \frac12m > 0\ ,
\end{align}
with $m>0$. This includes the case where the mass is much larger than the magnitudes of the Coulomb VEVs. In this phase, the FD-vertex simplifies as in \eqref{eq:red-FD} since one of the Young diagrams is empty.

One of our concerns would be to confirm the generic expression \eqref{eq:FD-vertex} for the FD-vertex, not just the reduced one \eqref{eq:red-FD}. Thus, it would be meaningful to compute the partition function in another phase, where we need to use the generic expression \eqref{eq:FD-vertex} for the FD-vertex, and show that it still reproduces the identical expected instanton partition function.

Another concern would be the consistency between the proposal in this paper and the one given in \cite{Kim:2024ufq}, where we have proposed the topological vertex formalism with an O7$^+$-plane not attached to any of the 5-branes. 
For this purpose, it would be interesting to consider the SU($2N$) gauge theory with a massless symmetric tensor and focus on the Coulomb branch point where the symmetric tensor can be Higgsed. After the Higgsing, the partition function should reduce to the one for the SO($2N$) gauge theory, which was computed in \cite{Kim:2024ufq}. Through this process, we would be able to see the relation between the two proposals. 

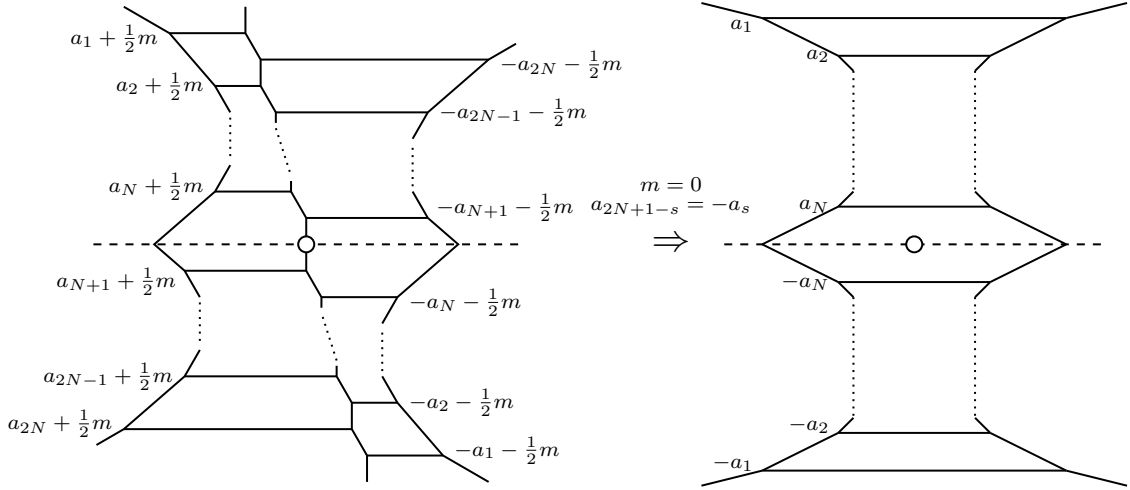
\begin{figure}
\centering
\begin{tikzpicture}
\begin{scriptsize}
\begin{scope}[xscale=0.4,yscale=0.7]
\draw[dashed] (-7,0) -- (7,0);
\draw (-6, 4.5) -- (-4.5, 4) -- (-3,3) -- (-2.5, 2.5);
\draw[dotted] (-2.5,2.4) -- (-2.5,1.6);
\draw (-2.5, 1.5) -- (-3, 1) -- (-5, 0) -- (-4, -0.5) -- (-3.5, -1);
\draw[dotted] (-3.5,-1.1) -- (-3.5,-1.9);
\draw (-3.5, -2) -- (-4,-2.5) -- (-6, -3.5) -- (-6.9, -3.8);
\node[anchor=east] at (-4.6,3.9) {$a_1+\frac12 m$};
\node[anchor=east] at (-3.1,3) {$a_2+\frac12 m$};
\node[anchor=east] at (-3.1,1.1) {$a_N+\frac12 m$};
\node[anchor=east] at (-4,-0.7) {$a_{N+1}+\frac12 m$};
\node[anchor=east] at (-4.1,-2.5) {$a_{2N-1}+\frac12 m$};
\node[anchor=east] at (-6.1,-3.4) {$a_{2N}+\frac12 m$};
\draw (6, -4.5) -- (4.5, -4) -- (3,-3) -- (2.5, -2.5);
\draw[dotted] (2.5,-2.4) -- (2.5,-1.6);
\draw (2.5, -1.5) -- (3, -1) -- (5, 0) -- (4, 0.5) -- (3.5, 1);
\draw[dotted] (3.5,1.1) -- (3.5,1.9);
\draw (3.5, 2) -- (4,2.5) -- (6,3.5) -- (6.9,3.8);
\node[anchor=west] at (4.6,-3.9) {$-a_1-\frac12 m$};
\node[anchor=west] at (3.1,-3) {$-a_2-\frac12 m$};
\node[anchor=west] at (3.1,-1.1) {$-a_N-\frac12 m$};
\node[anchor=west] at (4,0.7) {$-a_{N+1}-\frac12 m$};
\node[anchor=west] at (4.1,2.5) {$-a_{2N-1}-\frac12 m$};
\node[anchor=west] at (6.1,3.4) {$-a_{2N}-\frac12 m$};
\draw (-2,4.5) -- (-2,4) -- (-1.5,3.5) -- (-1.5,3) -- (-1,2.5) -- (-1,2.3);
\draw[dotted] (-1,2.2) -- (-0.5,1.3);
\draw (-0.5,1.2) -- (-0.5, 1) -- (0, 0.5) -- (0, 0);
\draw (2,-4.5) -- (2,-4) -- (1.5,-3.5) -- (1.5,-3) -- (1,-2.5) -- (1,-2.3);
\draw[dotted] (1,-2.2) -- (0.5,-1.3);
\draw (0.5,-1.2) -- (0.5, -1) -- (0, -0.5) -- (0, 0);
\draw (-4.5, 4) -- (-2, 4);
\draw (-3, 3) -- (-1.5, 3);
\draw (-3, 1) -- (-0.5, 1);
\draw (-4, -0.5) -- (0, -0.5);
\draw (-4, -2.5) -- (1, -2.5);
\draw (-6, -3.5) -- (1.5, -3.5);
\draw (4.5, -4) -- (2, -4);
\draw (3, -3) -- (1.5, -3);
\draw (3, -1) -- (0.5, -1);
\draw (4, 0.5) -- (0, 0.5);
\draw (4, 2.5) -- (-1, 2.5);
\draw (6, 3.5) -- (-1.5, 3.5);
\end{scope}
\draw[fill=white] (0,0) circle (3pt);
\begin{scope}[shift={(6,-1.5)}]

\draw[dashed] (-0.5, 1.5) -- (4.5, 1.5);

\node at (-1.2,1.5) {\Large $\Rightarrow$};
\node at (-1.2,2.3) {$m = 0$};
\node at (-1.2,2) {$a_{2N+1-s} = -a_s$};

\draw (0,1.5) -- (1,2);
\draw (1,2) -- (1.2, 2.2);
\draw (1,2) -- (3,2);
\draw (3,2) -- (4,1.5);
\draw (3,2) -- (2.8,2.2);
\draw (0,1.5) -- (1, 1);
\draw (1,1) -- (3,1);
\draw (3,1) -- (4,1.5);
\draw (1,1) -- (1.2,0.8);
\draw (3,1) -- (2.8,0.8);
\node[anchor=east] at (1,2) {$a_N$};
\node[anchor=east] at (1,1) {$-a_N$};

\node[anchor=east] at (1,-0.9) {$-a_2$};
\node[anchor=east] at (0,-1.4) {$-a_1$};
\draw (0,-1.5) -- (1,-1);
\draw (1.2,-0.8) -- (1,-1);
\draw (1,-1) -- (3,-1);
\draw (3,-1) -- (4,-1.5);
\draw (0,-1.5) -- (4,-1.5);
\draw (0,-1.5) -- (-0.8,-1.7);
\draw (4,-1.5) -- (4.8,-1.7);
\draw (3,-1) -- (2.8,-0.8);
\draw[dotted] (1.2,0.8) -- (1.2,-0.8);
\draw[dotted] (2.8,0.8) -- (2.8,-0.8);

\node[anchor=east] at (1,4) {$a_2$};
\node[anchor=east] at (0,4.4) {$a_1$};
\draw (1,4) -- (0,4.5);
\draw (0,4.5) -- (4,4.5);
\draw (4,4.5) -- (3,4);
\draw (1,4) -- (3,4);
\draw (0,4.5) -- (-0.8,4.7);
\draw (4,4.5) -- (4.8,4.7);
\draw (1,4) -- (1.2,3.8);
\draw (3,4) -- (2.8,3.8);
\draw[dotted] (1.2,3.8) -- (1.2,2.2);
\draw[dotted] (2.8,3.8) -- (2.8,2.2);

\draw[fill=white] (2, 1.5) circle (3pt);
\end{scope}

\end{scriptsize}
\end{tikzpicture}
\caption{5-brane web for the Higgsing from SU($2N$) gauge theory with a symmetric tensor. If we remove the central NS5-branes after the tuning, we obtain the 5-brane web diagram for the SO($2N$) gauge theory.}
\label{fig:Higgs-SUSO}
\end{figure}

Taking these two issues into account, we start with the 5-brane web diagram given on the left of Figure \ref{fig:Higgs-SUSO}, which realizes the SU($2N$) gauge theory with a symmetric tensor in the phase 
\begin{align}\label{eq:phase2}
a_1 + \frac12 m > -a_{2N} - \frac12 m > a_2 + \frac12 m > -a_{2N-1} - \frac12 m > \quad 
\cr
\cdots > a_N + \frac12 m > - a_{N+1} - \frac12 m > 0.
\end{align}
At a boundary of this phase, there exists a region 
\begin{align}\label{eq:tuning-gauge-Higgs}
m = 0, \quad a_{2N+1-s} = - a_{s},
\end{align}
where the Higgsing mentioned above occurs as depicted on the right of Figure \ref{fig:Higgs-SUSO}.

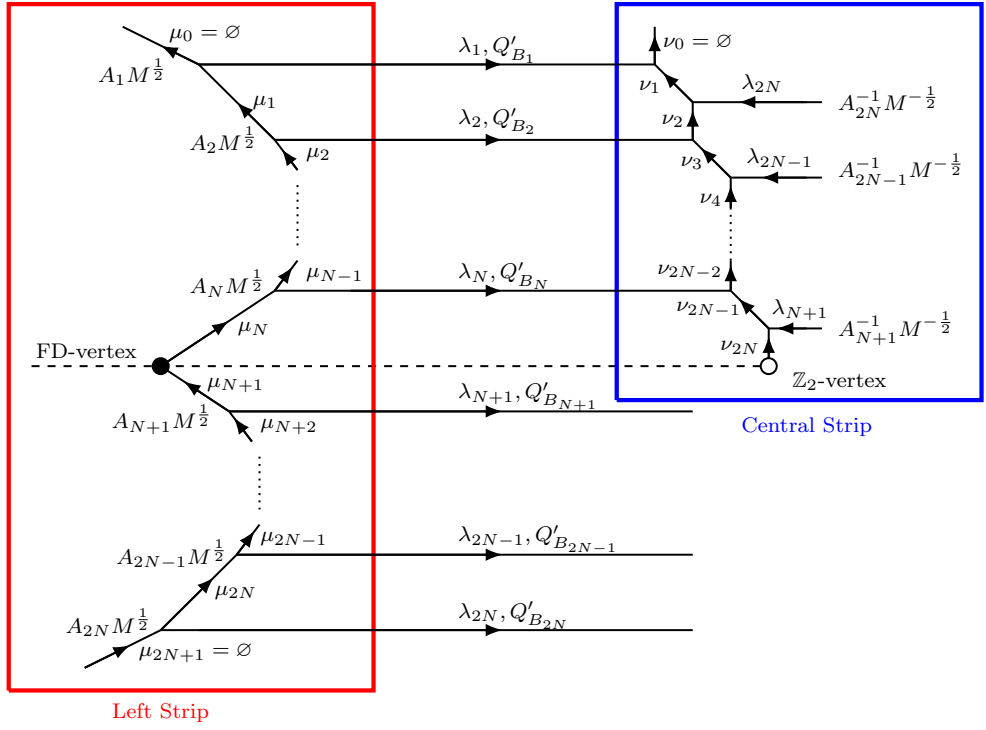
\begin{figure}
\centering
\begin{tikzpicture}
\begin{scriptsize}
\draw[dashed] (-0.7-5,0) -- (4,0);
\draw (-0.5-4, 4.5) -- (0.5-4, 4) -- (1.5-4,3) -- (1.8-4, 2.6) ;
\draw[dotted] (1.8-4,2.6-0.2) -- (1.8-4,1.4+0.2);
\draw (1.8-4, 1.4) -- (1.5-4, 1) -- (0-4, 0) -- (0.9-4, -0.6) -- (1.2-4, -1);
\draw[dotted] (1.3-4,-1-0.2) -- (1.3-4,-2.2+0.3);
\draw (1.3-4, -2.1) -- (1-4,-2.5) -- (0-4, -3.5) -- (-1-4, -4);
\draw[{Latex}-] (0-4, 4.25) -- (0.5-4, 4);
\draw[{Latex}-] (1-4, 3.5) -- (1.5-4, 3) ;
\draw[{Latex}-] (1.65-0.06-4, 2.8+0.08) -- (1.8-4, 2.6);
\draw[{Latex}-] (1.8-0.06-4, 1.4-0.08) -- (1.5-4, 1);
\draw[{Latex}-] (0.9-4, 0.6) -- (0-4, 0);
\draw[{Latex}-] (0.3-4, -0.2) -- (0.9-4, -0.6);
\draw[{Latex}-] (0.9+0.06-4, -0.6-0.08) -- (1.2-4, -1);
\draw[{Latex}-] (1.3-0.06-4, -2.1-0.08) -- (1-4, -2.5);
\draw[{Latex}-] (0.7-4, -2.8) -- (0-4, -3.5);
\draw[{Latex}-] (-0.4-4, -3.7) -- (-1-4, -4);
\node[anchor=east] at (0.2-4,3.9) {$A_1 M^{\frac{1}{2}}$};
\node[anchor=east] at (1.4-4,3) {$A_2 M^{\frac{1}{2}}$};
\node[anchor=east] at (1.5-4,1.1) {$A_N M^{\frac{1}{2}}$};
\node[anchor=east] at (0.8-4,-0.7) {$A_{N+1} M^{\frac{1}{2}}$};
\node[anchor=east] at (1-4,-2.5) {$A_{2N-1} M^{\frac{1}{2}}$};
\node[anchor=east] at (0-4,-3.4) {$A_{2N} M^{\frac{1}{2}}$};
\node[anchor=west] at (0-4,4.4) {$\mu_{0}=\varnothing$};
\node[anchor=west] at (1.1-4,3.5) {$\mu_1$};
\node[anchor=west] at (1.8-4,2.8) {$\mu_2$};
\node[anchor=west] at (1.8-4,1.2) {$\mu_{N-1}$};
\node[anchor=west] at (0.9-4,0.5) {$\mu_N$};
\node[anchor=west] at (0.5-4,-0.25) {$\mu_{N+1}$};
\node[anchor=west] at (1.2-4,-0.8) {$\mu_{N+2}$};
\node[anchor=west] at (1.2-4,-2.3) {$\mu_{2N-1}$};
\node[anchor=west] at (0.6-4,-3) {$\mu_{2N}$};
\node[anchor=west] at (-0.4-4,-3.8) {$\mu_{2N+1}=\varnothing$};
\draw[fill=black] (0-4,0) circle (3pt);
\node[anchor=east] at (-0.2-4,0.2) {FD-vertex};
\draw[ultra thick, red] (-2-4, -4.3) -- (-2-4, 4.8)-- (2.8-4, 4.8)-- (2.8-4, -4.3)-- (-2-4, -4.3);
\node at (1-5,-4.6){\textcolor{red}{Left Strip}};
\draw (2.5,4.5) -- (2.5,4) -- (3,3.5) -- (3,3) -- (3.5,2.5) -- (3.5,2.1);
\draw[dotted] (3.5,2.2) -- (3.5,1.4);
\draw (3.5,1.4) -- (3.5, 1) -- (4, 0.5) -- (4, 0);
\draw[{Latex}-] (2.5, 4.4) -- (2.5, 4);
\draw[{Latex}-] (2.6, 3.9) -- (3,3.5);
\draw[{Latex}-] (3, 3.4) -- (3,3);
\draw[{Latex}-] (3.1, 2.9) -- (3.5, 2.5);
\draw[{Latex}-] (3.5, 2.4) -- (3.5,2.1);
\draw[{Latex}-] (3.5, 1.35) -- (3.5, 1);
\draw[{Latex}-] (3.6, 0.9) -- (4, 0.5);
\draw[{Latex}-] (4, 0.4) -- (4, 0);
\draw (3,3.5) -- (4.7, 3.5);
\draw (3.5, 2.5) -- (4.7, 2.5);
\draw (4, 0.5) -- (4.7, 0.5);
\draw[{Latex}-] (3.6,3.5) -- (4.5, 3.5);
\draw[{Latex}-] (3.9, 2.5) -- (4.5, 2.5);
\draw[{Latex}-] (4.1, 0.5) -- (4.5, 0.5);
\node[anchor=west] at (4.8,3.5) {$A^{-1}_{2N}M^{-\frac{1}{2}}$};
\node[anchor=west] at (4.8,2.6) {$A^{-1}_{2N-1} M^{-\frac{1}{2}}$};
\node[anchor=west] at (4.8,0.5) {$A^{-1}_{N+1}M^{-\frac{1}{2}}$};
\node[anchor=west] at (2.5,4.3) {$\nu_{0}=\varnothing$};
\node[anchor=east] at (2.7,3.7) {$\nu_1$};
\node[anchor=east] at (3,3.25) {$\nu_2$};
\node[anchor=east] at (3.25,2.7) {$\nu_3$};
\node[anchor=east] at (3.5,2.2) {$\nu_4$};
\node[anchor=east] at (3.5,1.25) {$\nu_{2N-2}$};
\node[anchor=east] at (3.75,0.75) {$\nu_{2N-1}$};
\node[anchor=east] at (4,0.25) {$\nu_{2N}$};
\node[anchor=east] at (4.3,3.75) {$\lambda_{2N}$};
\node[anchor=east] at (4.7,2.75) {$\lambda_{2N-1}$};
\node[anchor=east] at (4.9,0.75) {$\lambda_{N+1}$};
\draw[fill=white] (4,0) circle (3pt);
\node[anchor=west] at (4.2,-0.2) {$\mathbb{Z}_2$-vertex};
\draw[ultra thick, blue] (2, -0.45) -- (2, 4.8)-- (6.8, 4.8)-- (6.8, -0.45)-- (2, -0.45);
\node at (4.5,-0.8){\textcolor{blue}{Central Strip}};
\draw (0.5-4, 4) -- (2.5, 4);
\draw (1.5-4, 3) -- (3, 3);
\draw (2.5-4, 1) -- (3.5, 1);
\draw (0.9-4, -0.6) -- (3, -0.6);
\draw (1-4, -2.5) -- (3, -2.5);
\draw (0-4, -3.5) -- (3, -3.5);
\draw[-{Latex}] (0.5-4, 4) -- (0.5, 4);
\draw[-{Latex}] (1.5-4, 3) -- (0.5, 3);
\draw[-{Latex}] (1.5-4, 1) -- (0.5, 1);
\draw[-{Latex}] (0.9-4, -0.6) -- (0.5, -0.6);
\draw[-{Latex}] (1-4, -2.5) -- (0.5, -2.5);
\draw[-{Latex}] (0.5-4, -3.5) -- (0.5, -3.5);
\node[anchor=west] at (3.8-4,4.2) {$\lambda_1, Q'_{B_1}$};
\node[anchor=west] at (3.8-4,3.25) {$\lambda_2, Q'_{B_2}$};
\node[anchor=west] at (3.8-4,1.2) {$\lambda_N, Q'_{B_N}$};
\node[anchor=west] at (3.8-4,-0.4) {$\lambda_{N+1}, Q'_{B_{N+1}}$};
\node[anchor=west] at (3.8-4,-2.3) {$\lambda_{2N-1}, Q'_{B_{2N-1}}$};
\node[anchor=west] at (3.8-4,-3.3) {$\lambda_{2N}, Q'_{B_{2N}}$};
\end{scriptsize}
\end{tikzpicture}
\caption{Assignment of Young diagrams and decomposition of the diagrams into strips for the 5d SU($N$) + 1{\bf Sym} in the phase from which 5d SO($2N$) is obtained.}
\label{fig:SUSO-top}
\end{figure}

First, we compute the topological string partition function based on the 5-brane web on the left of Figure \ref{fig:Higgs-SUSO}. As in the previous section, we consider the fundamental region and assign Young diagrams consistent with the $\mathbb{Z}_2$ action as in Figure \ref{fig:SUSO-top}. 
The K\"ahler parameters $Q'_{Bs}$ assigned to the horizontal edges are given in terms of the instanton factor $Q_I$ as
\begin{align}
Q'_{Bs} &= (-1)^{N+k} Q_I M^{-N+s-2} A_s^{N + k - 2 - s} 
\prod_{r=1}^{s-1} A_r 
\prod_{r=s+1}^{2N+1-s} A_r{}^{-1} 
\quad (s=1,\cdots N) , \
\cr
Q'_{Bs} &= (-1)^{N+k} Q_I M^{-N+s} A_s^{N + k + 2 - s} 
\prod_{r=2N+2-s}^{s-1} A_r
\prod_{r=s+1}^{2N} A_r^{-1}
\quad (s=N+1,\cdots 2N) . 
\cr
\end{align}

Analogous to the previous subsection, we compute the amplitude by computing the left strip and the central strip, respectively, and by gluing them as
\begin{align}\label{eq:Z-top-p2}
Z_{\text{top}}
&= \sum_{\mathbf{\lambda}} \prod_{s=1}^{2N} (-Q'_{B s}) ^{|\lambda_s|} 
\prod_{s=1}^N f_{\lambda_s}^{-N - k + 2  + s}
f_{\lambda_{2N+1-s}}^{N - k - 1  - s}
\times Z_{\text{left}} Z_{\text{central}} .
\end{align}
Again, by using our proposal, the central strip and the left strip are given, respectively, by
\begin{align}
Z_{\text{central}} 
&= \sum_{ \vec{\mathbf{\nu}} } 
\prod_{r=1}^{N} (-A_{r} A_{2N-r+1} M)^{|\nu_{2r-1}|} 
\cdot
\prod_{r=1}^{N-1} (-A_{2N-r+1}^{-1} A_{r+1}^{-1} M^{-1})^{|\nu_{2r}|} 
\cr
& \qquad 
\times
(A_{N+1}^{-1} M^{-\frac12})^{|\nu_{2N}|}
V^{\mathbb{Z}_2}_{\nu_{2N}} 
\prod_{s=1}^N C_{\nu_{2s-2} \nu_{2s-1}^T \lambda_s^T}
C_{\nu_{2s}^T \nu_{2s-1} \lambda_{2N-s+1}^T}\ , 
\cr
Z_{\text{left}} 
&= \sum_{ \vec{\mathbf{\mu}} } 
\prod_{r=1}^{N-1} (-A_{r} A_{r+1}^{-1})^{|\mu_r|} 
\cdot
\prod_{r=N+1}^{2N-1} (-A_{r} A_{r+1}^{-1})^{|\mu_{r+1}|} 
\cr
& \qquad \times (A_N M^{\frac12})^{|\mu_N|} (A_{N+1}^{-1} M^{-\frac12})^{|\mu_{N+1}|}
\cdot
\prod_{s=1}^{N-1} f_{\mu_s}^{-1}
\cdot
\prod_{s=N+2}^{2N} f_{\mu_s}^{-1}
\cr
& \qquad \times 
V^{\mathrm{FD}}_{\mu_{N+1} \mu_N} 
\prod_{s=1}^{N} C_{\mu_{s}^T \mu_{s-1} \lambda_s}
\prod_{s=N+1}^{2N} C_{\mu_{s+1}^T \mu_{s} \lambda_s}\ , 
\label{eq:LC-starting-p2}
\end{align}
where we define $\nu_0 = \mu_0 = \mu_{2N+1} = \varnothing$.

Computations analogous to the previous subsection show that the amplitudes for the central strip and the left strip can be written in the following form
\begin{align}\label{eq:LC-intermed-p2}
Z_{\text{central}} 
= & 
\prod_{r=1}^{2N} s_{\lambda_r} (q^{-\rho}) 
\times \prod_{r=1}^N R_{\lambda_{r} \lambda_{2N+1-r}} (A_r A_{2N+1-r} M) 
\cr
& \times 
\prod_{ 1 \le r < s \le N} 
R^{-1}_{\lambda_{r} \lambda_{s}^T} (A_r A_s^{-1})
R_{\lambda_{r} \lambda_{2N+1-s}} (A_r A_{2N+1-s} M)
\cr
& \times \prod_{N+1 \le r < s \le 2N} 
R^{-1}_{\lambda_{r} \lambda_{s}^T} (A_r A_s^{-1})
R_{ \lambda_{2N+1-r}^T \lambda_{s}^T} (A_{2N+1-r}^{-1} A_s^{-1} M^{-1}) \times \tilde{Z}_{\mathbb{Z}_2}
\cr
Z_{\text{left}} 
= & 
\prod_{r=1}^{2N} s_{\lambda_r^T} (q^{-\rho}) 
\!\!\!
\prod_{ 1 \le r < s \le N} \!\!\! R^{-1}_{\lambda_{r} \lambda_{s}^T} (A_r A_s^{-1})\!\!\!
\prod_{ N+1 \le r < s \le 2N} \!\!\!R^{-1}_{\lambda_{r} \lambda_{s}^T} (A_r A_s^{-1})
\times \tilde{Z}_{\text{FD}}, 
\end{align}
which are the counterparts of \eqref{eq:LC-intermed}. 
Here again, we have extracted the factors $\tilde{Z}_{\mathbb{Z}_2}$ and $\tilde{Z}_{\text{FD}}$, which  include the contributions from $\mathbb{Z}_2$ vertex and the FD-vertex, respectively. They are given in the operator formalism as
\begin{align}\label{eq:Z-CL-operator-p2}
\tilde{Z}_{\mathbb{Z}_2}
= & \bra{0}
\mathbb{O}_{\mathbb{Z}_2}
\prod_{r=1}^N \Gamma_-^{-1}(A_r M^{\frac12} q^{-\rho-\lambda_r})
\Gamma_- (A_{2N+1-r}^{-1} M^{-\frac12} q^{-\rho-\lambda_{2N+1-r}^T})
\ket{0}\ ,
\cr
\tilde{Z}_{\text{FD}}
= & \bra{0}
\prod_{s=N+1}^{2N} 
\Gamma_+ (A_s^{-1} M^{-\frac12} q^{-\rho - \lambda_s^T}) \, \cdot \mathbb{O}_{\mathrm{FD}}^- \mathbb{O}_{\mathrm{FD}}^+
\cdot \,
\prod_{r=1}^{N} 
\Gamma_- (A_r M^{\frac12} q^{-\rho - \lambda_r})
\ket{0}\ ,
\end{align}
analogous to \eqref{Z-CL-operator}. 
Unlike the previous case, we observe that the factor $\tilde{Z}_{\text{FD}}$ includes the operator $\mathbb{O}^-_{\text{FD}}$ in addition to $\mathbb{O}^+_{\text{FD}}$. 

The factor $\tilde{Z}_{\mathbb{Z}_2}$ is computed as 
\begin{align}\label{eq:ZZ2-p2-result}
\tilde{Z}_{\mathbb{Z}_2}
= &
\prod_{r=1}^N \prod_{s=1}^N 
R^{\frac12}_{\lambda_r \lambda_s}(A_r A_s M)
\times 
\prod_{r=N+1}^{2N} \prod_{s=N+1}^{2N} 
R^{\frac12}_{\lambda_r^T \lambda_s^T} (A_r^{-1} A_s^{-1} M^{-1})
\cr
& \times \prod_{r=1}^N \prod_{s=N+1}^{2N}
R^{-1}_{\lambda_r \lambda_s^T}(A_r A_s{}^{-1})
\end{align}
analogous to the case in the previous section.
As for the factor $\tilde{Z}_{\text{FD}}$ we use the following commutation relations sequentially, 
\begin{align}
\mathbb{O}_{\mathrm{FD}}^+ \Gamma_- (Q q^{-\rho-\mu}) 
&=   
\Gamma_- (Qq^{-\rho-\mu}) 
\mathbb{O}_{\mathrm{FD}}^+ 
\prod_{f=1}^4 \prod_{\ell = \pm 1}
R^{\frac12}_{\mu \varnothing} (\mathsf{M}_f^{\ell} Q) ,
\cr
\Gamma_+ (Q' q^{-\rho-\nu}) 
\mathbb{O}_{\mathrm{FD}}^-
&=    
\mathbb{O}_{\mathrm{FD}}^- 
\Gamma_+ (Q' q^{-\rho-\nu})
\prod_{f=1}^4 \prod_{\ell = \pm 1}
R^{\frac12}_{\nu \varnothing} (\mathsf{M}_f^{\ell} Q'),
\cr
\Gamma_+ (Q' q^{-\rho-\mu})
\Gamma_- (Q q^{-\rho-\nu}) 
&= 
\Gamma_- (Q q^{-\rho-\nu}) 
\Gamma_+ (Q' q^{-\rho-\mu})
R_{\mu\nu} (Q Q'),
\end{align}
with the identification $Q = A_r M^{\frac12}$, $Q' = A_s^{-1} M^{-\frac12}$.
This leads to
\begin{align}\label{eq:ZFD-p2-result}
\tilde{Z}_{\text{FD}}
= & 
\prod_{f=1}^4 \prod_{\ell = \pm 1} 
\left(
\prod_{r=1}^{N} R^{\frac12}_{\varnothing \lambda_r} (A_r M^{\frac12} \mathsf{M}_f^{\ell}) 
\times
\prod_{s=N+1}^{2N}  R^{\frac12}_{\varnothing \lambda_s^T} (A_s^{-1} M^{-\frac12} \mathsf{M}_f^{\ell})
\right) 
\cr
& \times \prod_{r=1}^{N} \prod_{s=N+1}^{2N} 
R^{-1}_{\lambda_r \lambda_s^T} (A_r A_s^{-1}). \
\end{align}

Combining all the results in \eqref{eq:Z-top-p2}, \eqref{eq:LC-intermed-p2}, \eqref{eq:ZZ2-p2-result}, and \eqref{eq:ZFD-p2-result}, we find that the topological string partition function in this phase is given by
\begin{align}
Z_{\text{top}}
&= \sum_{\mathbf{\lambda}} \prod_{s=1}^{2N} (-Q'_{B s}) ^{|\lambda_s|} 
\prod_{s=1}^N f_{\lambda_s}^{-N - k + 2  + s}
f_{\lambda_{2N+1-s}}^{N - k - 1  - s}
\cr
& \times \prod_{r=1}^{2N} s_{\lambda_r} (q^{-\rho}) s_{\lambda_r^T} (q^{-\rho})
\times \prod_{1 \le r < s \le 2N} R^{-2}_{\lambda_r \lambda_s^T}(A_r A_s{}^{-1})
\cr
& \times \prod_{\substack{r \ge 1, s \ge 1, 
\\ r+s \le 2N+1}} 
R^{\frac12}_{\lambda_r \lambda_s}(A_r A_s M)
\times  \prod_{\scriptstyle\substack{ r \le 2N, s \le 2N \\ r+s \ge 2N+2}} 
R^{\frac12}_{\lambda_r^T \lambda_s^T} (A_r^{-1} A_s^{-1} M^{-1})
\cr
& \times
\prod_{f=1}^4 \prod_{\ell = \pm 1} 
\left(
\prod_{r=1}^{N} R^{\frac12}_{\varnothing \lambda_r} (A_r M^{\frac12} \mathsf{M}_f^{\ell}) 
\times
\prod_{s=N+1}^{2N}  R^{\frac12}_{\varnothing \lambda_s^T} (A_s^{-1} M^{-\frac12} \mathsf{M}_f^{\ell})
\right). \
\end{align}

In the following, we check that this result is consistent with the result obtained in the previous subsection. 
The perturbative part obtained from this expression by taking the weak coupling limit is 
\begin{align}
\lim_{Q_I \to 0}Z_{\text{top}}
&= \prod_{1 \le r < s \le 2N} R^{-2}_{\varnothing\varnothing}(A_r A_s{}^{-1})
\cr
& \times \prod_{\substack{r \ge 1, s \ge 1, 
\\ r+s \le 2N+1}} 
R^{\frac12}_{\varnothing\varnothing}(A_r A_s M)
\times  \prod_{\substack{ r \le 2N, s \le 2N \\ r+s \ge 2N+2}} 
R^{\frac12}_{\varnothing\varnothing} (A_r^{-1} A_s^{-1} M^{-1})
\cr
& \times
\prod_{f=1}^4 \prod_{\ell = \pm 1} 
\left(
\prod_{r=1}^{N} R^{\frac12}_{\varnothing \lambda_r} (A_r M^{\frac12} \mathsf{M}_f^{\ell}) 
\times
\prod_{s=N+1}^{2N}  R^{\frac12}_{\varnothing \varnothing} (A_s^{-1} M^{-\frac12} \mathsf{M}_f^{\ell})
\right). \
\end{align}
This agrees with \eqref{eq:qto0limit} with $N \to 2N$ up to the replacement
\begin{align}
R_{\varnothing\varnothing} (Q) \to R_{\varnothing\varnothing} (Q^{-1}),
\end{align}
which corresponds to the flop transition.
Also, the instanton part 
\begin{align}
Z_{\text{inst}}
&= \frac{Z_{\text{top}}}{\lim\limits_{Q_I\to 0} Z_{\text{top}}}
\cr
&= \sum_{\mathbf{\lambda}} \prod_{s=1}^{2N} Q_{Bs}^{'|\lambda_s|} 
\prod_{s=1}^{N} 
f_{\lambda_s}^{-N - k + 2  + s}
f_{\lambda_{2N+1-s}}^{N - k - 1  - s}
\cr
& \qquad \times
\prod_{s=1}^{2N} N^{-1}_{\lambda_s \lambda_s}(1) 
\times
\prod_{1 \le s<t \le 2N} N^{-2}_{\lambda_t \lambda_s} (A_s A_t^{-1}) 
\cr
& \qquad 
\times \prod_{\substack{r \ge 1, s \ge 1, 
\\ r+s \le 2N+1}} 
N^{\frac12}_{\lambda_r^T \lambda_s}(A_r A_s M)
\times  \prod_{\substack{ r \le 2N, s \le 2N \\ r+s \ge 2N+2}} 
N^{\frac12}_{\lambda_r \lambda_s^T} (A_r^{-1} A_s^{-1} M^{-1})
\cr
& \qquad \times
\prod_{f=1}^4 \prod_{\ell = \pm 1} 
\left(
\prod_{r=1}^{N} N^{\frac12}_{\varnothing \lambda_r} (A_r M^{\frac12} \mathsf{M}_f^{\ell}) 
\times
\prod_{s=N+1}^{2N}  N^{\frac12}_{\varnothing \lambda_s^T} (A_s^{-1} M^{-\frac12} \mathsf{M}_f^{\ell})
\right) \
\end{align}
agrees with \eqref{eq:Z-inst} by replacing $N \to 2N$ after using the identity related to the flop transition:
\begin{align}
N_{\lambda \nu}(Q) = (-Q)^{|\lambda|+|\nu|} q^{\frac{\kappa(\lambda)}{2} - \frac{\kappa(\nu)}{2}} 
N_{\nu \lambda} (Q^{-1})\ .
\end{align}
This provides an additional consistency check for our proposal on the topological vertex formalism presented in section~\ref{sec:proposal}.

Finally, we discuss the limit where the Higgsing mentioned above occurs. For this purpose, it would be convenient to go back to the expression 
for the central strip given in \eqref{eq:LC-intermed-p2} and \eqref{eq:Z-CL-operator-p2}.
Tuning the parameters as
\begin{align}\label{eq:tuning-top-Higgs}
M \to 1 , \quad A_{2N+1-s} \to A_{s}^{-1}
\quad (s = 1,\cdots, N),
\end{align}
we find the first line in \eqref{eq:LC-intermed-p2}
vanishes unless the Young diagram satisfies $\lambda_{2N+1-r} = \lambda_{r}^T$
as 
\begin{align}
\prod_{r=1}^{2N} s_{\lambda_r^T} (q^{-\rho}) \times
\prod_{r=1}^N R_{\lambda_{r} \lambda_{2N+1-r}} (A_r A_{2N+1-r} M)
\to 
\tilde{M}(q)^{-N} \prod_{r=1}^N (-1)^{|\lambda_r|} \delta_{\lambda_{r}^T, \lambda_{2N+1-r}}\ , 
\end{align}
where $\tilde{M}(q):= \mathrm{PE} \big[ q/(1-q)^2\big]$ is the McMahon function.
Taking this into account, we find that the remaining factors in the central strip in \eqref{eq:LC-intermed-p2} all cancel out with each other in this limit, and the central strip reduces to
\begin{align}\label{eq:red-cent}
Z_{\text{central}} 
\to \tilde{M}(q)^{-N} \prod_{r=1}^N (-1)^{|\lambda_r|} \delta_{\lambda_{r}^T, \lambda_{2N+1-r}}
\quad (\text{as}~ M \to 1, \,\, A_{2N+1-s} \to A_{s}^{-1} ) \ .
\end{align}
This is consistent with the intuition based on Figure \ref{fig:Higgs-SUSO} that the central NS5-brane can be Higgsed away in this limit. 

In this limit, the partition function \eqref{eq:Z-top-p2} reduces to 
\begin{align}\label{eq:top-final}
Z_{\text{top}}
&= \tilde{M}(1)^{-N}
\sum_{\lambda_1, \cdots, \lambda_N} \prod_{s=1}^{N} ( - \tilde{Q}
_{B s} )^{|\lambda_s|} 
\prod_{s=1}^N f_{\lambda_s}^{-2N + 3  + 2s}
\times Z_{\text{left}} \biggl\vert_{\substack{
M=1, \,\, \\
A_{2N+1-s}=A_s^{-1}\\
\lambda_{2N+1-s} = \lambda_s^T  }}
 \ ,
\end{align}
where we denote
\begin{align}
\tilde{Q}_{B s} := Q'_{B s} Q'_{B_{2N+1-s}} \biggl\vert_{\substack{
M=1, \,\, \\
A_{2N+1-s}=A_s^{-1} }}
= Q_I^2 A_s^{2N-2-2s} \prod_{r=1}^{s-1} A_r^2\ . 
\end{align}

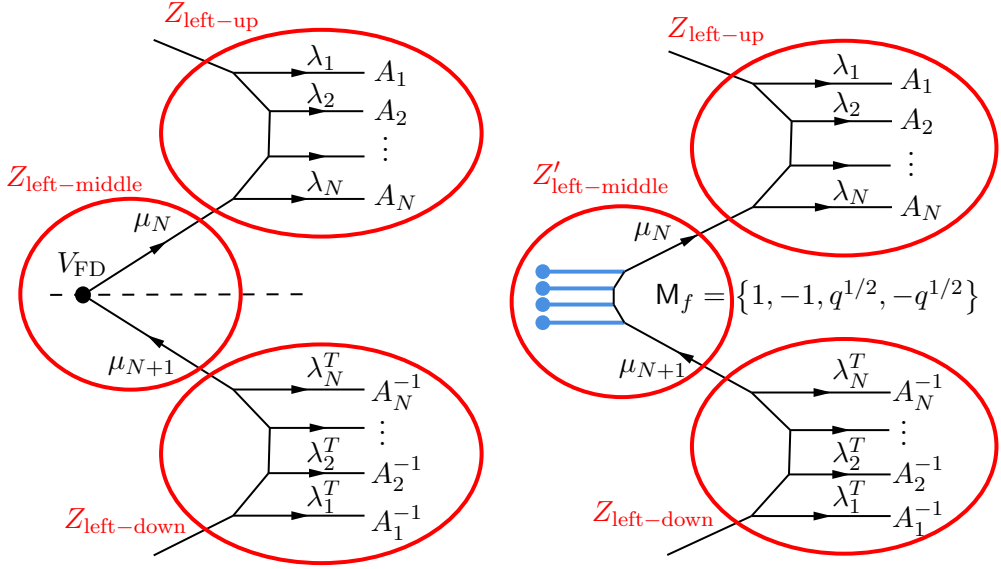
\begin{figure}
\centering
\begin{tikzpicture}[x=0.75pt,y=0.75pt,yscale=-0.75,xscale=0.75]
\draw  [dash pattern={on 4.5pt off 4.5pt}]  (31.53,228.45) -- (203.78,227.99) ;
\draw    (53.37,228.45) -- (153.36,164.09) ;
\draw [shift={(109.25,192.48)}, rotate = 147.23] [fill={rgb, 255:red, 0; green, 0; blue, 0 }  ][line width=0.08]  [draw opacity=0] (12,-3) -- (0,0) -- (12,3) -- cycle    ;
\draw    (153.36,164.09) -- (240.75,164.09) ;
\draw [shift={(204.06,164.09)}, rotate = 180] [fill={rgb, 255:red, 0; green, 0; blue, 0 }  ][line width=0.08]  [draw opacity=0] (12,-3) -- (0,0) -- (12,3) -- cycle    ;
\draw    (153.36,164.09) -- (176.89,135.58) ;
\draw    (176.89,135.58) -- (240.75,135.58) ;
\draw [shift={(215.82,135.58)}, rotate = 180] [fill={rgb, 255:red, 0; green, 0; blue, 0 }  ][line width=0.08]  [draw opacity=0] (12,-3) -- (0,0) -- (12,3) -- cycle    ;
\draw    (176.89,135.58) -- (177.73,105.24) ;
\draw    (177.73,105.24) -- (239.91,105.24) ;
\draw [shift={(215.82,105.24)}, rotate = 180] [fill={rgb, 255:red, 0; green, 0; blue, 0 }  ][line width=0.08]  [draw opacity=0] (12,-3) -- (0,0) -- (12,3) -- cycle    ;
\draw    (177.73,105.24) -- (153.36,79.49) ;
\draw    (153.36,79.49) -- (240.75,79.49) ;
\draw [shift={(204.06,79.49)}, rotate = 180] [fill={rgb, 255:red, 0; green, 0; blue, 0 }  ][line width=0.08]  [draw opacity=0] (12,-3) -- (0,0) -- (12,3) -- cycle    ;
\draw    (100.43,57.42) -- (153.36,79.49) ;
\draw    (53.37,228.45) -- (153.36,291.9) ;
\draw [shift={(95.77,255.35)}, rotate = 32.4] [fill={rgb, 255:red, 0; green, 0; blue, 0 }  ][line width=0.08]  [draw opacity=0] (12,-3) -- (0,0) -- (12,3) -- cycle    ;
\draw    (153.36,291.9) -- (240.75,291.9) ;
\draw [shift={(204.06,291.9)}, rotate = 180] [fill={rgb, 255:red, 0; green, 0; blue, 0 }  ][line width=0.08]  [draw opacity=0] (12,-3) -- (0,0) -- (12,3) -- cycle    ;
\draw    (177.73,317.65) -- (153.36,291.9) ;
\draw    (177.73,317.65) -- (240.75,317.65) ;
\draw [shift={(216.24,317.65)}, rotate = 180] [fill={rgb, 255:red, 0; green, 0; blue, 0 }  ][line width=0.08]  [draw opacity=0] (12,-3) -- (0,0) -- (12,3) -- cycle    ;
\draw    (176.89,347.99) -- (177.73,317.65) ;
\draw    (176.89,347.99) -- (240.75,347.99) ;
\draw [shift={(215.82,347.99)}, rotate = 180] [fill={rgb, 255:red, 0; green, 0; blue, 0 }  ][line width=0.08]  [draw opacity=0] (12,-3) -- (0,0) -- (12,3) -- cycle    ;
\draw    (153.36,376.5) -- (176.89,347.99) ;
\draw    (153.36,376.5) -- (240.75,376.5) ;
\draw [shift={(204.06,376.5)}, rotate = 180] [fill={rgb, 255:red, 0; green, 0; blue, 0 }  ][line width=0.08]  [draw opacity=0] (12,-3) -- (0,0) -- (12,3) -- cycle    ;
\draw    (100.43,403.16) -- (153.36,376.5) ;
\draw  [fill={rgb, 255:red, 0; green, 0; blue, 0 }  ,fill opacity=1 ] (48.75,228.45) .. controls (48.75,225.66) and (50.82,223.4) .. (53.37,223.4) .. controls (55.93,223.4) and (57.99,225.66) .. (57.99,228.45) .. controls (57.99,231.25) and (55.93,233.51) .. (53.37,233.51) .. controls (50.82,233.51) and (48.75,231.25) .. (48.75,228.45) -- cycle ;
\draw    (413.72,212.84) -- (499.2,169.61) ;
\draw [shift={(462.7,188.06)}, rotate = 153.17] [fill={rgb, 255:red, 0; green, 0; blue, 0 }  ][line width=0.08]  [draw opacity=0] (12,-3) -- (0,0) -- (12,3) -- cycle    ;
\draw    (499.2,169.61) -- (591.81,169.61) ;
\draw [shift={(552.5,169.61)}, rotate = 180] [fill={rgb, 255:red, 0; green, 0; blue, 0 }  ][line width=0.08]  [draw opacity=0] (12,-3) -- (0,0) -- (12,3) -- cycle    ;
\draw    (499.2,169.61) -- (524.13,141.68) ;
\draw    (524.13,141.68) -- (590.92,141.68) ;
\draw [shift={(564.52,141.68)}, rotate = 180] [fill={rgb, 255:red, 0; green, 0; blue, 0 }  ][line width=0.08]  [draw opacity=0] (12,-3) -- (0,0) -- (12,3) -- cycle    ;
\draw    (524.13,141.68) -- (525.02,111.96) ;
\draw    (525.02,111.96) -- (591.81,111.96) ;
\draw [shift={(565.41,111.96)}, rotate = 180] [fill={rgb, 255:red, 0; green, 0; blue, 0 }  ][line width=0.08]  [draw opacity=0] (12,-3) -- (0,0) -- (12,3) -- cycle    ;
\draw    (525.02,111.96) -- (499.2,86.74) ;
\draw    (499.2,86.74) -- (591.81,86.74) ;
\draw [shift={(552.5,86.74)}, rotate = 180] [fill={rgb, 255:red, 0; green, 0; blue, 0 }  ][line width=0.08]  [draw opacity=0] (12,-3) -- (0,0) -- (12,3) -- cycle    ;
\draw    (443.1,65.12) -- (499.2,86.74) ;
\draw    (413.72,247.07) -- (498.31,293.91) ;
\draw [shift={(448.14,266.13)}, rotate = 28.97] [fill={rgb, 255:red, 0; green, 0; blue, 0 }  ][line width=0.08]  [draw opacity=0] (12,-3) -- (0,0) -- (12,3) -- cycle    ;
\draw    (498.31,293.91) -- (590.92,293.91) ;
\draw [shift={(551.61,293.91)}, rotate = 180] [fill={rgb, 255:red, 0; green, 0; blue, 0 }  ][line width=0.08]  [draw opacity=0] (12,-3) -- (0,0) -- (12,3) -- cycle    ;
\draw    (524.13,319.13) -- (498.31,293.91) ;
\draw    (524.13,319.13) -- (590.03,319.13) ;
\draw [shift={(564.08,319.13)}, rotate = 180] [fill={rgb, 255:red, 0; green, 0; blue, 0 }  ][line width=0.08]  [draw opacity=0] (12,-3) -- (0,0) -- (12,3) -- cycle    ;
\draw    (523.24,348.85) -- (524.13,319.13) ;
\draw    (523.24,348.85) -- (590.92,348.85) ;
\draw [shift={(564.08,348.85)}, rotate = 180] [fill={rgb, 255:red, 0; green, 0; blue, 0 }  ][line width=0.08]  [draw opacity=0] (12,-3) -- (0,0) -- (12,3) -- cycle    ;
\draw    (498.31,376.77) -- (523.24,348.85) ;
\draw    (498.31,376.77) -- (590.92,376.77) ;
\draw [shift={(551.61,376.77)}, rotate = 180] [fill={rgb, 255:red, 0; green, 0; blue, 0 }  ][line width=0.08]  [draw opacity=0] (12,-3) -- (0,0) -- (12,3) -- cycle    ;
\draw    (442.21,402.89) -- (498.31,376.77) ;
\draw [color={rgb, 255:red, 74; green, 144; blue, 226 }  ,draw opacity=1 ][line width=1.5]    (359.4,212.84) -- (413.72,212.84) ;
\draw    (413.72,212.84) -- (406.59,224.1) ;
\draw [color={rgb, 255:red, 74; green, 144; blue, 226 }  ,draw opacity=1 ][line width=1.5]    (359.4,224.1) -- (406.59,224.1) ;
\draw [color={rgb, 255:red, 74; green, 144; blue, 226 }  ,draw opacity=1 ][line width=1.5]    (359.4,234.91) -- (406.59,234.91) ;
\draw [color={rgb, 255:red, 74; green, 144; blue, 226 }  ,draw opacity=1 ][line width=1.5]    (359.4,247.07) -- (413.72,247.07) ;
\draw    (406.59,224.1) -- (406.59,234.91) ;
\draw    (406.59,234.91) -- (413.72,247.07) ;
\draw  [draw opacity=0][fill={rgb, 255:red, 74; green, 144; blue, 226 }  ,fill opacity=1 ] (354.5,212.84) .. controls (354.5,210.1) and (356.69,207.89) .. (359.4,207.89) .. controls (362.1,207.89) and (364.29,210.1) .. (364.29,212.84) .. controls (364.29,215.58) and (362.1,217.79) .. (359.4,217.79) .. controls (356.69,217.79) and (354.5,215.58) .. (354.5,212.84) -- cycle ;
\draw  [draw opacity=0][fill={rgb, 255:red, 74; green, 144; blue, 226 }  ,fill opacity=1 ] (354.5,224.1) .. controls (354.5,221.36) and (356.69,219.15) .. (359.4,219.15) .. controls (362.1,219.15) and (364.29,221.36) .. (364.29,224.1) .. controls (364.29,226.84) and (362.1,229.05) .. (359.4,229.05) .. controls (356.69,229.05) and (354.5,226.84) .. (354.5,224.1) -- cycle ;
\draw  [draw opacity=0][fill={rgb, 255:red, 74; green, 144; blue, 226 }  ,fill opacity=1 ] (354.5,234.91) .. controls (354.5,232.17) and (356.69,229.95) .. (359.4,229.95) .. controls (362.1,229.95) and (364.29,232.17) .. (364.29,234.91) .. controls (364.29,237.64) and (362.1,239.86) .. (359.4,239.86) .. controls (356.69,239.86) and (354.5,237.64) .. (354.5,234.91) -- cycle ;
\draw  [draw opacity=0][fill={rgb, 255:red, 74; green, 144; blue, 226 }  ,fill opacity=1 ] (354.5,247.07) .. controls (354.5,244.33) and (356.69,242.11) .. (359.4,242.11) .. controls (362.1,242.11) and (364.29,244.33) .. (364.29,247.07) .. controls (364.29,249.8) and (362.1,252.02) .. (359.4,252.02) .. controls (356.69,252.02) and (354.5,249.8) .. (354.5,247.07) -- cycle ;

\draw (212,120) [color=red, line width=1.5] ellipse (80pt and 52pt);
\draw (85,228) [color=red, line width=1.5] ellipse (55pt and 48pt);
\draw (212,335) [color=red, line width=1.5] ellipse (80pt and 55pt);
\draw (560,130) [color=red, line width=1.5] ellipse (76pt and 54pt);
\draw (412,233) [color=red, line width=1.5] ellipse (55pt and 48pt);
\draw (560,330) [color=red, line width=1.5] ellipse (76pt and 54pt);

\draw (200,143) node [anchor=north west][inner sep=0.75pt]    {$\lambda _{N}$};
\draw (200,84) node [anchor=north west][inner sep=0.75pt]    {$\lambda _{2}$};
\draw (200,59) node [anchor=north west][inner sep=0.75pt]    {$\lambda _{1}$};
\draw (34.01,197.79) node [anchor=north west][inner sep=0.75pt]    {$V_{\mathrm{FD}}$};
\draw (245,153) node [anchor=north west][inner sep=0.75pt]    {$A_{N}$};
\draw (245,96) node [anchor=north west][inner sep=0.75pt]    {$A_{2}$};
\draw (245,70) node [anchor=north west][inner sep=0.75pt]    {$A_{1}$};
\draw (250,110) node [anchor=north west][inner sep=0.75pt]    {$\vdots $};
\draw (200,265) node [anchor=north west][inner sep=0.75pt]    {$\lambda _{N}^{T}$};
\draw (200,322) node [anchor=north west][inner sep=0.75pt]    {$\lambda _{2}^{T}$};
\draw (200,351) node [anchor=north west][inner sep=0.75pt]    {$\lambda _{1}^{T}$};
\draw (243,280) node [anchor=north west][inner sep=0.75pt]    {$A_{N}^{-1}$};
\draw (243,335) node [anchor=north west][inner sep=0.75pt]    {$A_{2}^{-1}$};
\draw (243,367.77) node [anchor=north west][inner sep=0.75pt]    {$A_{1}^{-1}$};
\draw (248,300) node [anchor=north west][inner sep=0.75pt]    {$\vdots $};
\draw (85,172) node [anchor=north west][inner sep=0.75pt]    {$\mu _{N}$};
\draw (68,268) node [anchor=north west][inner sep=0.75pt]    {$\mu _{N+1}$};
\draw (105,30) node [anchor=north west][inner sep=0.75pt]  [color=red]  {$Z_{\mathrm{left-up}}$};
\draw (0,140) node [anchor=north west][inner sep=0.75pt]  [color=red]  {$Z_{\mathrm{left-middle}}$};
\draw (38.71,369.73) node [anchor=north west][inner sep=0.75pt]  [color=red]  {$Z_{\mathrm{left-down}}$};
\draw (550,149) node [anchor=north west][inner sep=0.75pt]    {$\lambda _{N}$};
\draw (550,90) node [anchor=north west][inner sep=0.75pt]    {$\lambda _{2}$};
\draw (550,65) node [anchor=north west][inner sep=0.75pt]    {$\lambda _{1}$};
\draw (595,160) node [anchor=north west][inner sep=0.75pt]    {$A_{N}$};
\draw (595,102) node [anchor=north west][inner sep=0.75pt]    {$A_{2}$};
\draw (595,75) node [anchor=north west][inner sep=0.75pt]    {$A_{1}$};
\draw (600,120) node [anchor=north west][inner sep=0.75pt]    {$\vdots $};
\draw (550,266) node [anchor=north west][inner sep=0.75pt]    {$\lambda _{N}^{T}$};
\draw (550,322) node [anchor=north west][inner sep=0.75pt]    {$\lambda _{2}^{T}$};
\draw (550,351) node [anchor=north west][inner sep=0.75pt]    {$\lambda _{1}^{T}$};
\draw (590,280) node [anchor=north west][inner sep=0.75pt]    {$A_{N}^{-1}$};
\draw (590,335) node [anchor=north west][inner sep=0.75pt]    {$A_{2}^{-1}$};
\draw (590,365) node [anchor=north west][inner sep=0.75pt]    {$A_{1}^{-1}$};
\draw (595,300) node [anchor=north west][inner sep=0.75pt]    {$\vdots $};
\draw (418,178) node [anchor=north west][inner sep=0.75pt]    {$\mu _{N}$};
\draw (408,270) node [anchor=north west][inner sep=0.75pt]    {$\mu _{N+1}$};
\draw (432,215) node [anchor=north west][inner sep=0.75pt]    {$\mathsf{M}_{f} =\left\{1,-1,q^{1/2} ,-q^{1/2}\right\}$};
\draw (440,40) node [anchor=north west][inner sep=0.75pt]  [color=red]  {${Z}{_{\mathrm{left-up}}}$};
\draw (350,137) node [anchor=north west][inner sep=0.75pt]  [color=red]  {$Z'_{\mathrm{left-middle}}$};
\draw (390,365) node [anchor=north west][inner sep=0.75pt]  [color=red]  {$Z_{\mathrm{left-down}}$};
\end{tikzpicture}
\caption{Comparison between two formalisms for computing partition function for 5d SO($2N$) gauge theory. Left: Formalism based on the FD vertex proposed in this paper. Right: Formalism based on four virtual D7-branes.}
\label{fig:FD-vs-4D7}
\end{figure}

Going back to the expression for the left strip in \eqref{eq:LC-starting-p2}, we find
\begin{align}\label{eq:red-left-p2}
Z_{\text{left}} \biggl\vert_{\substack{
M=1, \,\, \\
A_{2N+1-s}=A_s^{-1}\\
\lambda_{2N+1-s} = \lambda_s^T  }}
&= \sum_{ \mu_N,\, \mu_{N+1} } 
Z_{\text{upper-left}} \,
Z_{\text{middle-left}} \,
Z_{\text{lower-left}} ,
\cr
Z_{\text{upper-left}} & 
:= \sum_{\mu_1, \cdots , \mu_{N-1}} \prod_{r=1}^{N-1} (-A_{r} A_{r+1}^{-1})^{|\mu_r|} 
 f_{\mu_r}^{-1}
 \times \prod_{s=1}^{N} C_{\mu_{s}^T \mu_{s-1} \lambda_s}
\\
Z_{\text{middle-left}} &
:= 
(-A_N)^{|\mu_N|+|\mu_{N+1}|}
V^{\mathrm{FD}}_{\mu_{N+1} \mu_N} \,
\cr
Z_{\text{lower-left}} & 
:=
\sum_{\mu_{N+1}, \cdots , \mu_{2N}}  
\prod_{r=N+2}^{2N} (-A_{r} A_{r-1}^{-1})^{|\mu_{r}|}  f_{\mu_r}^{-1}
\times \prod_{s=N+1}^{2N} C_{\mu_{s+1}^T \mu_{s} \lambda_{s}^T} \biggl\vert_{\substack{
A_{2N+1-s}=A_s^{-1}\\
\lambda_{2N+1-s} = \lambda_s^T  }}\ .
\nonumber
\end{align}
Here, we split the expression for the contribution from the left strip into three parts: the upper half contribution $Z_{\text{upper-left}}$, the lower half contribution $Z_{\text{lower-left}}$, and the remaining part $Z_{\text{middle-left}}$ including the FD vertex with the adjacent edge factors, as depicted on the left of Figure \ref{fig:FD-vs-4D7}.
In other words, we find that the Nekrasov's instanton partition function for 5d $\mathcal{N}=1$ pure SO($2N$) gauge theory, whose 5-brane web diagram is given to the right of Figure \ref{fig:Higgs-SUSO}, should be computed according to the left in Figure~\ref{fig:FD-vs-4D7}, by using the FD-vertex. This indicates that our proposal for the FD vertex should also apply to cases where an O7$^+$-plane is not attached to any 5-branes as in configurations for SO gauge theories.

This proposal should be compared with the proposal in~\cite{Kim:2024ufq}, based on which the partition function for the SO($2N$) theory is given in the form
\begin{align}\label{eq:top-SO}
Z_{\text{SO}}
&= 
\sum_{\lambda_1, \cdots, \lambda_N} \prod_{s=1}^{N} ( - \tilde{Q}
_{B s} )^{|\lambda_s|} 
\prod_{s=1}^N f_{\lambda_s}^{-2N + 3  + 2s}
Z_{\text{strip}}\ ,
\cr
Z_{\text{strip}} 
&= 
\sum_{ \mu_N, \mu_{N+1} } 
Z_{\text{upper-left}} \,
Z'_{\text{middle-left}} \,
Z_{\text{lower-left}}. \
\end{align}
Here, the first line has a structure identical to \eqref{eq:top-final} up to the McMahon function, which is an overall factor that does not depend on the Coulomb moduli and should be discarded. Also, $Z_{\text{upper-left}}$ and $Z_{\text{lower-left}}$ in the second line are exactly the same as those defined in \eqref{eq:red-left-p2}, respectively. Meanwhile, the remaining part $Z'_{\text{middle-left}}$ is given by
\begin{align}\label{eq:middle-prime}
Z'_{\text{middle-left}} :=  
&\sum_{\nu_1, \nu_2, \nu_3}
C_{\nu_1 \mu_N^T \varnothing}
C_{\nu_2 \nu_1^T \varnothing}
C_{\nu_3 \nu_2^T \varnothing}
C_{\mu_{N+1} \mu_3^T \varnothing}
\cr
& \times 
(-A_N \mathsf{M}_1^{-1})^{|\mu_N|} 
\left( \prod_{r=1}^3 (- \mathsf{M}_r \mathsf{M}_{r+1}^{-1})^{|\nu_r|}
f_{\nu_r}
\right) (-\mathsf{M}_4 A_N)^{|\mu_{N+1}|} \ .
\end{align}
This ``middle part'' corresponds to the contribution from the four virtual D7 branes, which generate four virtual flavor D5 branes due to the Hanany-Witten transition, as depicted on the right in Figure~\ref{fig:FD-vs-4D7}. 

In this comparison, we find that $Z'_{\text{left-middle}}$ in \eqref{eq:middle-prime}, which is the contribution from the four virtual D7 branes proposed in~\cite{Kim:2024ufq}, is replaced by $Z_{\text{left-middle}}$ in \eqref{eq:red-left-p2}, which is the contribution from the FD-vertex, in this paper. Then, from \eqref{eq:LC-intermed-p2} and \eqref{eq:ZFD-p2-result}, we find that the left strip reduces to 
\begin{align}\label{eq:left-red-p2}
Z_{\text{left}} \biggl\vert_{\substack{M=1, \,\,\\ A_{2N+1-s}=A_s^{-1}\\
\lambda_{2N+1-s} = \lambda_s^T  }
}
= &\prod_{r=1}^{N} s_{\lambda_r} (q^{-\rho}) s_{\lambda_r^T} (q^{-\rho}) 
\times \prod_{ 1 \le r < s \le N} 
R^{-2}_{\lambda_r \lambda_s^T} (A_r A_s^{-1})
\cr
& 
\times \prod_{r=1}^N \prod_{s=1}^N 
R^{-1}_{\lambda_r \lambda_s^T} (A_r A_s)
\times
\prod_{f=1}^4 \prod_{\ell = \pm 1}
\prod_{r=1}^{N} R_{\varnothing \lambda_r} (A_r \mathsf{M}_f^{\ell}) \ . \
\end{align}
 This is identical to $Z_{\text{strip}}$ in~\eqref{eq:top-SO}, which had been already computed in \cite{Kim:2024ufq}, after changing the convention for the indices for the Coulomb moduli~$A_s$. Thus, the partition functions for SO($2N$) gauge theory computed in~\eqref{eq:top-final} and that in~\eqref{eq:top-SO} are identical to each other. This indicates that our proposal in this paper is consistent with the proposal given in~\cite{Kim:2024ufq}.

However, we do not necessarily claim that $Z_{\text{left-middle}}$ and $Z'_{\text{left-middle}}$ are identical to each other for arbitrary Young diagrams $\mu_{N}$ and $\mu_{N+1}$. In fact, if we replace the FD-vertex by the four virtual D7 branes in the case of SU($N$) gauge theory with a symmetric tensor with generic mass, we will not obtain the correct answer. Indeed, the second line in~\eqref{eq:Nek-SU+Sym-final} implies that the correct expression, which we have confirmed to be reproduced by our proposal using the FD-vertex, has half of the contribution of eight flavors rather than the usual contribution of four flavors. However, if we tune the parameters as in~\eqref{eq:tuning-gauge-Higgs} or \eqref{eq:tuning-top-Higgs}, half of the contribution of eight flavors reduces to the usual contribution of four flavors. In this way, when $Z_{\text{left-middle}}$ or $Z'_{\text{left-middle}}$ are embedded into the 5-brane web diagrams as in Figure~\ref{fig:FD-vs-4D7}, the partition functions computed using these two become identical to each other. This discussion implies that we can use either FD-vertex or four virtual flavor branes if none of the 5-branes is attached to the O7$^+$-plane as in~\cite{Kim:2024ufq}. In contrast, we should use the FD-vertex when one of the 5-branes is attached to the O7$^+$-plane as in this paper.

\bigskip
\section{Conclusion}\label{sec:conclusion}
In this paper, we proposed a novel (unrefined) topological vertex formulation for five-dimensional $\mathcal{N}=1$ SU($N$) supersymmetric gauge theories with a hypermultiplet in the symmetric tensor multiplet. The brane construction is realized using Type IIB brane webs with an O7$^+$-plane to which an NS5-brane is attached, enabling the realization of the brane configurations corresponding to a symmetric hypermultiplet. Building upon our previous work~\cite{Kim:2024ufq}, where we developed a topological vertex formalism for brane setups with an O7$^+$-plane where no NS5-brane is attached, and computed the partition function for $\mathcal{N}=1$ SO($2N$) gauge theories, we generalized the formulation to include a symmetric hypermultiplet in supersymmetric SU($N$)$_k$ gauge theory at Chern-Simons level~$k$.

We reformulated the conventional Nekrasov instanton partition function using the new Nekrasov factor. This reformulation provides a more intuitive interpretation of the instanton partition function as the decomposition of the instanton partition function as a product of contributions from W-bosons, a bifundamental hypermultiplet, and eight virtual flavors.

For the topological vertex formulation associated with a 5-brane web with an O7$^+$-plane where an NS5-brane is attached, we introduced two new vertices: the $\mathbb{Z}_2$-vertex \eqref{eq:Z2-vertex} and the FD-vertex \eqref{eq:FD-vertex}. Both are defined in the operator formulation analogous to the O-vertex~\cite{Hayashi:2020hhb, Nawata:2021dlk}. The $\mathbb{Z}_2$-vertex constitutes an explicit realization of the $\mathbb{Z}_2$ orbifold action, which we identified by disentangling two distinct operators present in the O-vertex. The FD-vertex also arises from the O-vertex by identifying the contributions for four frozen (virtual) D5-branes that intersect with the monodromy cut of the O7$^+$-plane. Expressed as the operator formalism, the contribution from the FD vertex in the partition function clearly shows that from frozen D5-branes with particular tuned masses. We interpret these operators as representing the $\mathbb{Z}_2$ orbifold action and the contributions of four frozen (virtual) D7-branes, together corresponding to the structure O7$^+\sim \mathbb{Z}_2+4\mathrm{D7s}$.

We tested our proposal through various checks, which include consistency with possible phases between the Coulomb branch parameters and the mass parameter of the symmetric hypermultiplet. In particular, in the regime where the mass of symmetric hypermultiplet is taken to zero, we observe that the anticipated Higgs branch where the middle NS5-brane decouples along the transverse directions to the 5-brane web, while the Coulomb branch parameters recombine to form continuous connections. This corresponds to the Higgsing of the SU$(2N)_k$ gauge theory with a symmetric hypermultiplet to the SO$(2N)$ gauge theory, in agreement with our expectations.

As a generalization of \cite{Kim:2024ufq}, our proposal enables us to compute the partition function for the SU$(2N)_k$ gauge theory with a symmetric hypermultiplet based on a 5-brane web with an NS5-brane attached to an O7$^+$-plane. It is worth noting that the FD-vertex is applicable for the 5-brane configuration without an NS5-brane attached to an O7$^+$-plane. In other words, one can directly apply the FD-vertex to compute the SO$(2N)$ gauge theory. 

Equipped with this new topological vertex formulation associated with an O7$^+$-plane, one immediate application is to compute~\cite{workinprogress} the partition function for local $\mathbb{P}^2$ with an adjoint matter proposed by \cite{Bhardwaj:2019jtr}, whose brane configuration involves an O7$^+$-plane where an NS5-brane is attached \cite{Kim:2020hhh}. Another direction to pursue is to generalize topological vertex formulation to brane configuration with an O7$^-$-plane, which leads to Sp($N$) gauge theories and SU$(N)_k$ gauge theories with an antisymmetric hypermultiplet.

\acknowledgments
SK is supported by National Natural Science Foundation of China (NSFC) grant No. 12250610188. XL is supported by NSFC grant No.11501470, No.11426187, No.11791240561, the Fundamental Research Funds for the Central Universities 2682021ZTPY043 and partially supported by NSFC grant No. 11671328. Especially, XL would like to thank Bohui Chen, An-min Li, Guosong Zhao for their constant support and also thank all the friends met in different conferences. FY is supported by start-up research grant RK21121 from Huzhou University. RZ is supported by NSFC No. 12105198.
\bigskip

\appendix

\section{Identities of Nekrasov factor}\label{sec:App}
We adopt the notations and conventions as presented in~\cite{Kim:2024ufq}. For example, we use the Greek letters $\lambda, \nu, \cdots$ to denote Young diagrams, where
$\lambda = \big((\lambda)_1, (\lambda)_2, \cdots \big)$. Here,  $(\lambda)_{i}$ denotes the number of boxes in the $i$-th row of $\lambda$. A box at the $i$-th row and the $j$-th column is denoted as $(i,j)$. The transpose of a Young diagram $\lambda$ is denoted by  $\lambda^T$. The total number of boxes of a given Young diagram $\lambda$ is given by $|\lambda| = \sum_{(i,j)\in\lambda} 1 = \sum_{i} (\lambda)_i= |\lambda^T|$. 
 When denoting Young diagrams collectively, we use 
$\vec{\lambda} = (\lambda_1, \lambda_2, \cdots )$. For clarity, we point out that $\lambda_i$ should not be confused with  $(\lambda)_i$.

The Nekrasov factor used in this paper is defined by
\begin{equation}
N_{\lambda\nu}(Q):=\prod_{(i,j)\in \lambda}\lt(1-Qq^{1-i-j+(\lambda)_i+(\nu^T)_j}\rt)\prod_{(m,n)\in \nu}\lt(1-Qq^{m+n-(\lambda^T)_n-(\nu)_m-1}\rt)\ ,
\label{def:Nekra}
\end{equation}
and satisfies the identity
\begin{align}
N_{\lambda\nu}(Q)
&=
N_{\nu^T \lambda^T}(Q)
=
(-Q)^{|\lambda|+|\nu|}q^{\frac{1}{2}\kappa(\lambda)-\frac{1}{2}\kappa(\nu)} N_{\nu\lambda} \lt( Q^{-1} \rt)\ , 
\label{Nekra-convert-formu}
\end{align}
where $\kappa(\lambda):=2\sum_{(i,j)\in\lambda}(j-i)$, which corresponds to the second Casimir of the representation of the unitary group given by the Young diagram $\lambda$. When $Q=1$, the Nekrasov factor is related to the Schur function as
\begin{align}
N^{-1}_{\lambda\lambda}(1)
&=
(-1)^{|\lambda|}s_\lambda(q^{-\rho})s_{\lambda^T}(q^{-\rho})\ .
\label{id:Schur-Nekra}
\end{align}

We introduce the R-factor, defined by \begin{align}\label{eq:PE}
R_{\mu \nu} (Q) 
:= \prod_{i=1}^{\infty} \prod_{j=1}^{\infty}
\lt( 1 - Q q^{i+j-1-(\mu)_i-(\nu)_j} \rt)\ .
\end{align}
The Nekrasov factor is then related to the R-factor as
\begin{align} \label{eq:R-definition}
R_{\mu \nu} (Q) 
= \mathrm{PE} \left[ - \frac{q}{(1-q)^2} Q \right] N_{\mu^T \nu} (Q)\ ,
\end{align}
where PE is the plethystic exponential defined as
\begin{align}
\mathrm{PE}\big[f(x_i)\big] = \exp\bigg( \sum_{n=1}^{\infty} \frac1n f(x_i^n)\bigg)  \ . 
\end{align}
The Cauchy identities are then expressed as follows:
\begin{align}
\sum_\lambda &~ s_{\lambda/\mu}(Q_1 q^{-\rho-\sigma})s_{\lambda/\nu}(Q_2 q^{-\rho-\tau})
\cr
&~
= R^{-1}_{\sigma \tau}(Q_1 Q_2) \sum_\eta s_{\nu/\eta}(Q_1 q^{-\rho-\sigma})s_{\mu/\eta}(Q_2 q^{-\rho-\tau})\ ,\label{Schur-nor-id-spec}
\\
\sum_\lambda &~ s_{\lambda/\mu^T}(Q_1 q^{-\rho-\sigma})s_{\lambda^T/\nu}(Q_2 q^{-\rho-\tau})
\cr
&~=R_{\sigma \tau}(Q_1 Q_2)\sum_\eta s_{\nu^T/\eta}(Q_1 q^{-\rho-\sigma})s_{\mu/\eta^T}(Q_2 q^{-\rho-\tau})\ .
\label{Schur-twist-id-spec}
\end{align}

We also introduce a \emph{new Nekrasov factor} $\tilde{n}_{\lambda\nu}$, defined as 
\begin{align}
\tilde{n}_{\lambda\nu}(a) 
:= &\prod_{(i,j)\in \lambda}
\mathrm{sh}\Big(a + \hbar \big(1-i-j+(\lambda)_i+(\nu^T)_j \big)\Big)
\cr
& \quad \times
\prod_{(i,j)\in \nu}
\mathrm{sh} \Big(a + \hbar \big(i+j-1-(\lambda^T)_j-(\nu)_i \big)\Big)\ ,
\end{align}
where $\mathrm{sh}(z) =e^{z/2}- e^{-z/2} $.
It is related to the original Nekrasov factor $N_{\lambda\nu}$ as 
\begin{align}\label{eq:nN}
\tilde{n}_{\lambda\nu}(a) 
= 
A^{-\frac12 (|\lambda|+|\nu|)} q^{-\frac14 (\kappa(\lambda) - \kappa(\nu))}
N_{\lambda\nu}(A)\ ,
\end{align}
with $A=e^{-a}$ and $q = e^{-\hbar}$.
This satisfies the following identities:
\begin{align}\label{eq:sym-new-nek}
\tilde{n}_{\lambda\nu}(a)
= \tilde{n}_{\nu^T \lambda^T}(a)
= (-1)^{|\lambda|+|\nu|} \tilde{n}_{\nu\lambda}(-a)
= (-1)^{|\lambda|+|\nu|} \tilde{n}_{\lambda^T \nu^T}(-a)\ .
\end{align}
When one of the Young diagrams is empty, it is given by
\begin{align}\label{eq:Nek-empty}
\tilde{n}_{\lambda^T \varnothing}(a) 
= &\prod_{(i,j)\in \lambda^T}
2\, \sinh \frac12 \Big(a + \hbar (1-i-j+(\lambda^T)_i )\Big)
\cr
= &\prod_{(i,j)\in \lambda}
2\, \sinh \frac12\Big(a + \hbar (i-j )\Big)\ ,
\end{align}
which then leads to yet another identities:
\begin{align}
\tilde{n}_{\lambda^T\varnothing}(a)
= \tilde{n}_{\varnothing \lambda}(a)
= (-1)^{|\lambda|} \tilde{n}_{\lambda \varnothing}(-a)
= (-1)^{|\lambda|} \tilde{n}_{\varnothing\lambda^T}(-a)\ .
\end{align}
\bigskip

In the following, we discuss identities regarding the
Nekrasov factor defined in \eqref{def:Nekra} and the factor defined by
\begin{align}\label{eq:def-of-Y-main}
Y_\lambda(Q)
:=\frac{\displaystyle\prod_{x\in \mathfrak{A}(\lambda)}\Big(1-Q \chi_x \Big)}{\displaystyle\prod_{x\in \mathfrak{R}(\lambda)}\Big(1-Q \chi_x \Big)}, \qquad 
Y_\varnothing(Q) = 1-Q \ ,
\end{align}
with $\chi_x = q^{i-j}$. 
Here, $\mathfrak{A}(\lambda)$ denotes the set of boxes that can be added to $\lambda$, while $\mathfrak{R}(\lambda)$ denotes the set of boxes that can be removed from $\lambda$. 
This $Y_\lambda(Q)$ satisfies the following recursion relation
\begin{align}\label{eq:recY}
\frac{Y_{\lambda+x}(Q)}{Y_{\lambda}(Q)}=\frac{(1-Q\chi_x q)(1-Q\chi_x q^{-1})}{(1-Q\chi_x)^2}\ ,
\end{align}
where $\lambda+x$ denotes a Young diagram that is obtained by adding one box at $x$ to $\lambda$. 
Also, with this $Y_\lambda(Q)$, it is straightforward to see that the Nekrasov factors satisfy the following recursion relations:
\begin{align}
\frac{N_{\lambda_1^T (\lambda_2+x)}(Q)}{N_{\lambda_1^T \lambda_2}(Q)}
&= Y_{\lambda_1}(Q \chi_x)\ ,&
\frac{N_{(\lambda_1+x)^T\lambda_2}(Q)}{N_{\lambda_1{}^T \lambda_2}(Q)} 
&= Y_{\lambda_2}(Q \chi_x)\ .
\label{Nek-rec-Y}
\end{align}

Here, we also introduce 
\begin{align}
\tilde{Y}_{\lambda}(Q) 
&:= (1-Q) \prod_{x \in \lambda} \frac{(1- Q \chi_x q)(1- Q \chi_x q^{-1})}{(1- Q \chi_x)^2},
\cr
\tilde{N}_{\lambda\mu}(Q)
&:= \prod_{x \in \lambda} (1-Q\chi_x)
\times\!
\prod_{y \in \mu} (1-Q\chi_y)
\times\!
\prod_{\substack{x \in \lambda \\{y \in \mu}}} 
\frac{(1- Q \chi_x \chi_y q)(1- Q \chi_x \chi_y q^{-1})}{(1- Q \chi_x \chi_y)^2}.
\end{align}
In the following, we show that they are identical to $Y_{\lambda}(Q)$ and $N_{\lambda^T\mu}(Q)$, respectively,
by considering the recursion relation.

First, we discuss $\tilde{Y}_{\lambda}(Q)$ and $Y_{\lambda}(Q)$. It is obvious to see the agreement of the initial condition:
\begin{align}
\tilde{Y}_{\varnothing}(Q) = 1 - Q = Y_{\varnothing}(Q)\ .
\end{align}
Also, it is obvious from the definition of $\tilde{Y}_{\lambda}(Q)$ that it satisfies the identical recursion relation as $Y_{\lambda}(Q)$:
\begin{align}
\frac{\tilde{Y}_{\lambda+x}(Q)}{\tilde{Y}_{\lambda}(Q)}
= \frac{(1- Q \chi_x q)(1- Q \chi_x q^{-1})}{(1- Q \chi_x)^2}
= \frac{Y_{\lambda+x}(Q)}{Y_{\lambda}(Q)},
\end{align}
Thus, we find that they are identical to each other:
\begin{align}
\tilde{Y}_{\lambda}(Q) = Y_{\lambda}(Q)\ .
\end{align}

Next, we go on to $\tilde{N}_{\lambda\mu}(Q)$. Again, the agreement of the initial condition is obvious:
\begin{align}
\tilde{N}_{\varnothing\varnothing}(Q) 
= 1 
= N_{\varnothing\varnothing}(Q)\ .
\end{align}
Here, note that it can be rewritten in terms of $\tilde{Y}_{\lambda}(Q)$ in the following two ways:
\begin{align}
\tilde{N}_{\lambda\mu}(Q)
&= \prod_{x \in \lambda}(1-Q\chi_x) \times 
\prod_{y \in \mu}\tilde{Y}_{\lambda}(Q\chi_y)\ 
\cr
&= \prod_{y \in \mu}(1-Q\chi_y) \times 
\prod_{x \in \lambda}\tilde{Y}_{\mu}(Q\chi_x)\ .
\end{align}
From these expressions, we show that $\tilde{N}_{\lambda\mu}(Q)$ satisfies the identical recursion relation as $N_{\lambda\mu}(Q)$ 
\begin{align}
\frac{\tilde{N}_{\lambda (\mu+y)}(Q)}{\tilde{N}_{\lambda \mu}(Q)}
= \tilde{Y}_{\lambda}(Q\chi_y)
= Y_{\lambda}(Q\chi_y)
= \frac{N_{\lambda^T(\mu+y)}(Q)}{N_{\lambda^T \mu}(Q)}\ ,
\cr
\frac{\tilde{N}_{(\lambda+x) \mu}(Q)}{\tilde{N}_{\lambda \mu}(Q)}
= \tilde{Y}_{\mu}(Q\chi_x)
= Y_{\mu}(Q\chi_x)
= \frac{N_{\lambda^T(\mu+x)}(Q)}{N_{\lambda^T \mu}(Q)} \ .
\end{align}
Thus, we have shown 
\begin{align}
\tilde{N}_{\lambda \mu}(Q) = N_{\lambda^T \mu}(Q)\ .
\end{align}

Rewriting this in terms of the new Nekrasov factor $\tilde{n}_{\lambda\mu}(a)$, we find
\begin{align}\label{eq:key-id-n}
\tilde{n}_{\lambda^T\mu}(a)
=
&\prod_{(i,j) \in \lambda} \mathrm{sh}(a+(i-j)\hbar)
\times
\prod_{(m,n) \in \mu} \mathrm{sh}(a+(m-n)\hbar)
\cr
&
\times
\prod_{\substack{(i,j) \in \lambda \\{(m,n) \in \mu}}} 
\frac{\mathrm{sh} (a+(i-j+m-n \pm 1)\hbar )}{\mathrm{sh}^2 (a+(i-j+m-n)\hbar)}\ ,
\end{align}
where we use the notation $\mathrm{sh} (x \pm y) := \mathrm{sh}(x+y) \mathrm{sh}(x-y)$.
This identity plays an important role in this paper.

\section{Comparison to the real topological vertex}\label{sec:App-RTV}

In this appendix, we compare the real topological vertex studied in \cite{Walcher:2007qp, Krefl:2009md, Krefl:2009mw} with the ``topological vertex with orientifold'' studied in this paper and in \cite{Kim:2024ufq, Kim:2017jqn}. 
On the one hand, the real topological vertex is formulated for toric Calabi-Yau 3-folds with $\mathbb{Z}_2$ orientifold, often accompanied by a D-brane on top of the orientifold plane. On the other hand, ``topological vertex with orientifold'' is formulated for 5-brane web diagrams with an O5-plane or an O7-plane. Since orientifolds are introduced in both cases, it may look natural to guess that they are equivalent formalisms. However, we will see that they are actually different, implying that introducing an orientifold plane in the 5-brane web diagram does not correspond to introducing an orientifold plane in the toric Calabi-Yau 3-folds.

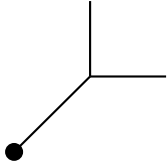
\begin{figure}[ht]
\centering
\begin{tikzpicture}[
    line/.style = {thick},
    dot/.style = {circle, fill=black, inner sep=0pt, minimum size=2pt},
    label/.style = {font=\small}
]

\draw[line] (0,0) -- (-1,-1); 
\draw[line] (0,0) -- (1,0); 
\draw[line] (0,0) -- (0,1); 

\draw[fill=black] (-1,-1) circle (3pt);

\end{tikzpicture}
\caption{``Half'' of the resolved conifold. }
\label{fig:resolved-conifold}
\end{figure}

To see the difference, we consider the resolved conifold as an example depicted in Figure~\ref{fig:resolved-conifold}.  %
\noindent From the viewpoint of the real topological vertex, this is the conifold with involution of ``type 1'' \cite{Krefl:2009mw}. This type 1 involution is related to the $\mathbb{Z}_2$ action that rotates this diagram by $\pi$ radians. In this setup, the $S^2$ in the resolved conifold is quotiented by $\mathbb{Z}_2$ action to give $\mathbb{R}P^2$, and the whole geometry gives morally ``half'' of the resolved conifold. It is found that the partition function computed with the real topological vertex is given by \cite{Krefl:2009mw, Bouchard:2004iu}
\begin{align}\label{eq:rcon-real}
 Z = \exp \left( - \frac12 \sum_{k=1}^{\infty} \frac{Q^k}{k(q^{\frac{k}{2}}-q^{-\frac{k}{2}})^2} \pm \sum_{\genfrac{}{}{0pt}{}{k=1}{k\text{:odd}}}^{\infty} \frac{Q^{\frac{k}{2}}}{k(q^{\frac{k}{2}}-q^{-\frac{k}{2}})} \right).
\end{align}
Here, the parameter $q$ is related to the Omega deformation parameters $\hbar = \epsilon_1 = -\epsilon_2$ as $q=e^{-\hbar}$. Two more types of involution, related to the reflection, are also studied in \cite{Krefl:2009mw}. All these turn out to give the identical result \eqref{eq:rcon-real}.

Probably, the most similar setup in the 5-brane web diagram is given by using the same shape of 5-brane web diagram with O7$^+$-plane located at the black node in Figure~\ref{fig:resolved-conifold}. In this case, we assign the $\mathbb{Z}_2$-vertex at the black node, while we do not need to use the FD-vertex in this setup. The partition function for this setup is given by
\begin{align}\label{eq:rcon-or}
Z = \exp \left( - \frac12 \sum_{k=1}^{\infty} \frac{Q^k}{k(q^{\frac{k}{2}}-q^{-\frac{k}{2}})^2} \right).
\end{align}
We observe that the second term in \eqref{eq:rcon-real} is missing in \eqref{eq:rcon-or}. This implies that the real topological vertex does not cover the geometries corresponding to the 5-brane web diagrams with an O7-plane and vice versa.


As expected from the conifold transition between the resolved conifold and the deformed conifold, the closed topological string partition function on the resolved conifold (with no orientifold)
\begin{align}\label{eq:rcon}
Z = \exp \left( - \sum_{k=1}^{\infty} \frac{Q^k}{k(q^{\frac{k}{2}}-q^{-\frac{k}{2}})^2}  \right)
\end{align}
agrees with the partition function for 3d SU($N$) Chern-Simons theory on $S^3$ \cite{Gopakumar:1998ki}. 
Here, this $N$ is related to the Omega deformation parameter $\hbar$ as 
\begin{align}\label{eq:h-N}
\hbar~ \sim~1/N\ .
\end{align}

The corresponding statement for the orientifold case, in the context of the real topological vertex, is that the closed topological string partition function \eqref{eq:rcon-real} on a resolved conifold with involution agrees with 3d SO($2N$)/Sp($N$) Chern-Simons theory on $S^3$ \cite{Sinha:2000ap}. The plus sign corresponds to Sp group while the minus sign corresponds to SO group. We note that the other two types of involution studied in \cite{Krefl:2009mw} correspond to 3d SU($N$) Chern-Simons theory on $S^3/\mathbb{Z}_2$, which also give compatible results.

In contrast, the partition function \eqref{eq:rcon-or} does not admit a straightforward interpretation in terms of 3d Chern-Simons theory. Notably, we observe that it constitutes precisely ``half'' of the 3d SU($N$) Chern-Simons partition function on $S^3$ given in \eqref{eq:rcon}. This also implies that 5-brane web diagrams with an orientifold plane studied in the context of ``topological vertex with orientifold'' should not be interpreted as the orientifold of the Calabi-Yau 3-folds.


``The topological vertex with orientifolds'' is formulated based on 5-brane web diagrams with an O-plane. The D5-branes and an O-plane in the 5-brane web diagram fill 5d spacetime, and realize 5d $\mathcal{N}=1$ gauge theories. This class of gauge theories includes SO($2N'$)/Sp($N'$) gauge theories or SU($N'$) gauge theory with a hypermultiplet in the symmetric/anti-symmetric representation. Note that the rank of the gauge group $N'$ here should be distinguished from that which appears in \eqref{eq:h-N}. This $N'$ is determined by the number of D5-branes in the 5-brane web diagram and has nothing to do with the Omega deformation parameters.

On the contrary, it is less obvious to understand the gauge theory interpretation for the real topological string in general. If we would like to interpret it in the context of the Gopakumar-Vafa invariant \cite{Gopakumar:1998ii, Gopakumar:1998jq}, it is natural to assume that the $\mathbb{Z}_2$ action on the resolved conifold also acts on the 4d spacetime as $(-1,-1,1,1)$ as discussed in \cite{Sinha:2000ap, Walcher:2007qp, Piazzalunga:2014waa}. This introduces a codimension-2 defect to a 4d supersymmetric gauge theory at the $\mathbb{Z}_2$ invariant locus. The first term in \eqref{eq:rcon-real} corresponds to BPS particles propagating in 4d spacetime, while the second term corresponds to the BPS particles moving only inside the defect \cite{Sinha:2000ap}. If we uplift to M-theory, it is expected to compute the ``twisted Nekrasov partition function'' of 5d supersymmetric gauge theory \cite{Hayashi:2015uka}. However, the interpretation along this direction is not entirely understood. 

Instead, it would also be possible to consider a connection with 4d supersymmetric gauge theories without defect by introducing an O6-plane and $N$ D6-branes filling the 4d spacetime and the $S^3$ in the deformed conifold as discussed in \cite{Sinha:2000ap}. This setup, realizing 3d SO($2N$)/Sp($N$) Chern-Simons theory of $S^3$, can be seen as 4d $\mathcal{N}=1$ SO($2N$)/Sp($N$) Yang-Mills theory from the viewpoint of 4d spacetime. This time, $N$ is the same one given in \eqref{eq:h-N}. In this context, the closed topological string partition function on the resolved conifold with the involution, which is obtained after the conifold transition, can be used to understand the glueball superpotential, which appears in the IR confining phase. The relation with the Nekrasov partition function is less manifest in this picture.

Thus, we claim that the real topological vertex and the topological vertex with orientifolds each study different, and most likely disjoint, classes of 5d gauge theories.

\bigskip
\bibliographystyle{JHEP}
\bibliography{ref}
\end{document}